\documentclass[12pt]{article}
\usepackage{rangecite}
\usepackage{epsfig}
\usepackage{rotating}

\newcommand{\gsim}{~{}_{\textstyle\sim}^{\textstyle >}~}

\def\OEE{\Omega_{\eta+\eta'}}

\def\kpnn{K^+\rightarrow\pi^+\nu\bar\nu}

\def\klpn{K_{\rm L}\rightarrow\pi^0\nu\bar\nu}
\def\bmumu{B_s\rightarrow\mu\bar\mu}
\def\bnunu{B\rightarrow X_s\nu\bar\nu}
\def\klee{K_{\rm L}\rightarrow\pi^0 e^+e^-}

\newcommand{\imlt}{\IM\lambda_t}
\newcommand{\relt}{\RE\lambda_t}
\newcommand{\relc}{\RE\lambda_c}

\newcommand{\mz}{M_{\rm Z}}
\newcommand{\bea}{\begin{eqnarray}}
\newcommand{\eea}{\end{eqnarray}}
\newcommand{\bd}{\begin{displaymath}}
\newcommand{\ed}{\end{displaymath}}
\newcommand{\be}{\begin{equation}}
\newcommand{\ee}{\end{equation}}
\newcommand{\ord}{{\cal O}}

\newcommand{\eps}{\varepsilon}
\newcommand{\epe}{\varepsilon'/\varepsilon}
\newcommand{\mt}{m_{\rm t}}
\newcommand{\mc}{m_{\rm c}}
\newcommand{\ms}{m_{\rm s}}
\newcommand{\md}{m_{\rm d}}
\newcommand{\mw}{M_{\rm W}}
\newcommand{\gev}{\, {\rm GeV}}
\newcommand{\mev}{\, {\rm MeV}}

\newcommand{\bsi}{B_6^{(1/2)}}
\newcommand{\bei}{B_8^{(3/2)}}

\newcommand{\Lms}{\Lambda_{\overline{\rm MS}}}

\newcommand{\RE}{{\rm Re}}
\newcommand{\IM}{{\rm Im}}
\def\pl#1#2#3{{ Phys.~Lett. }{\bf B#1~}(19#2)~#3}

\def\prl#1#2#3{{ Phys. Rev. Lett. }{\bf #1~}(19#2)~#3}

\def\pr#1#2#3{{ Phys. Rev. }{\bf D#1~}(19#2)~#3}
\def\np#1#2#3{{ Nucl. Phys. }{\bf B#1~}(19#2)~#3}

\newcommand{\vcb}{|V_{cb}|}
\newcommand{\vtd}{|V_{td}|}
\newcommand{\vts}{|V_{ts}|}
\newcommand{\vub}{|V_{ub}/V_{cb}|}

\newcommand{\beq}{\begin{equation}}
\newcommand{\eeq}{\end{equation}}

\def\con{\ifmmode \hbox{\bf*} \else{\bf*}\fi}   
\def\scon{\ifmmode \hbox{\footnotesize\rm\bf*} \else{\footnotesize\rm\bf*}\fi}

\def\0#1{\relax\ifmmode\mathaccent"7017{#1}
        \else\accent23#1\relax\fi}              

\def\eps{\varepsilon}

\newcommand{\W}{{\scriptscriptstyle W}}
\newcommand{\HH}{{\scriptscriptstyle H}}

\newcommand{\sq}{{\tilde q}}
\newcommand{\supq}{{\tilde u}}
\newcommand{\sdown}{{\tilde d}}
\newcommand{\stp}{{\tilde t}}
\newcommand{\tht}{{\theta_{\tilde t}}}
\newcommand{\tm}{{\tilde m}}
\newcommand{\tM}{{\tilde M}}

\newcommand{\tV}{{\tilde V}}
\newcommand{\tU}{{\tilde U}}
\newcommand{\half}{\frac{1}{2}}

\begin{document}


\author{ {\large\bf A.J.~Buras${}^{1}$, P. Gambino${}^{2}$,
    M. Gorbahn${}^{1}$,}  \\
  {\large\bf S. J\"ager${}^{1}$
    and L. Silvestrini${}^{1,3}$} \\
  \ \\
  {\small ${}^{1}$ Physik Department,
    Technische Universit\"at M\"unchen,} \\
  {\small D-85748 Garching, Germany} \\
  {\small ${}^{2}$ Theory Division, CERN, CH-1211 Geneva 23, 
    Switzerland} \\
  {\small ${}^{3}$ Dipartimento di Fisica, Universit\`a di Roma
    ``La Sapienza'' and} \\
  {\small INFN, Sezione di Roma, P.le A. Moro, I-00185, Roma, Italy}
}
\date{}
\title{
{\normalsize\sf
\rightline{TUM-HEP-382/00}
}
\bigskip
{\LARGE\bf
$\epe$ and Rare K and \\
B Decays in the MSSM 
}}

\maketitle
\thispagestyle{empty}

\phantom{xxx} \vspace{-9mm}

\begin{abstract}
  We analyze the CP violating ratio $\epe$ and rare K and B decays in
  the MSSM with minimal flavour and CP violation, including NLO QCD
  corrections and imposing constraints on the supersymmetric
  parameters coming from $\varepsilon$, $B_{d,s}^0-\bar B_{d,s}^0$
  mixings, $B\to X_s \gamma$, $\Delta\varrho$ in the electroweak
  precision studies and from the lower bound on the neutral Higgs
  mass. We provide a compendium of phenomenologically relevant
  formulae in the MSSM.  Denoting by $T(Q)$ the MSSM prediction for a
  given quantity normalized to the Standard Model result we find the
  ranges: $0.53\le T(\epe)\le 1.07$, $0.65\le T(\kpnn)\le 1.02$,
  $0.41\le T(\klpn)\le 1.03$, $0.48\le T(\klee)\le 1.10$, $0.73\le
  T(\bnunu)\le 1.34$ and $0.68\le T(\bmumu)\le 1.53$.  We point out
  that these ranges will be considerably reduced when the lower bounds
  on the neutral Higgs mass and $\tan\beta$ improve. Some contour
  plots illustrate the dependences of the quantities above on the
  relevant supersymmetric parameters. As a byproduct of this work we
  update our previous analysis of $\epe$ in the SM and find in NDR
  $\epe=(9.2^{+6.8}_{-4.0})$, a value $15\%$ higher than in our 1999
  analysis.
 \end{abstract}

\newpage
\setcounter{page}{1}
\setcounter{footnote}{0}  

\section{Introduction}
\setcounter{equation}{0}

One of the present central issues of particle physics phenomenology is
the question whether the size of the experimentally observed
direct CP violation in $K_L\to\pi\pi$ decays can be described within
the Standard Model (SM).

The most recent experimental results for the ratio $\epe$,
which measures the size of the direct CP violation in $K_L\to\pi\pi$
relative to the indirect one, read:
\begin{equation}\label{eprime1}
\RE(\varepsilon'/\varepsilon) =\left\{ \begin{array}{ll}
(28.0 \pm 4.1)\cdot 10^{-4} & {\rm (KTeV)}~\cite{KTEV} \\
(14.0 \pm 4.3)\cdot 10^{-4} &{\rm (NA48)}~ \cite{NA48}.
\end{array} \right.
\end{equation}
Together with the older NA31 measurement ($(23 \pm 7)\cdot 10^{-4}$)
\cite{barr:93}, these data confidently establish direct 
CP violation in nature and taking also the
E731 result $((7.4 \pm 5.9)\cdot 10^{-4})$ \cite{gibbons:93}
into account one finds
the grand average 
\be
\RE(\epe) = (19.2\pm 2.4)\cdot 10^{-4} \quad \mathrm{with} \quad 
\chi^2/ndf=11.1/5
\label{ga}
\ee
close to the NA31 result. 

There are different opinions whether the grand average in (\ref{ga})
can be accommodated within the SM. We list the results from various
groups in table \ref{tab:31738}.  The labels (MC) and (S) in the
second column stand for two error estimates: ``Monte Carlo'' and
``Scanning'' respectively.  The result from the Munich group given
here is an update of the analysis in \cite{EP99} done in the present
paper. Similarly the result of the Rome group is the most recent
estimate given in \cite{CM00}.  We do not list the values of the
non-perturbative parameters $\bsi$ and $\bei$ used in different
analyses as the comparison of these values could be misleading.  The
reason is that in certain papers these factors are $\ms-$independent
and in other papers they depend on $\ms$. Typically $\bsi=0.8-1.6$ and
$\bei=0.6-1.2$. Exceptions are the analyses in \cite{Narison} and
\cite{Prades} where $\bsi$ in the ballpark of 3 can be found.

In \cite{EP99,CM00,ROMA99} $\epe$ has been found to be typically a
factor of 2 below the data and the KTeV result in (\ref{eprime1})
could only be accommodated if all relevant parameters were chosen
simultaneously close to their extreme values.  On the other hand the
NA48 result is essentially compatible with \cite{EP99,CM00,ROMA99}
within experimental and theoretical errors.  Higher values of $\epe$
than in \cite{EP99,CM00,ROMA99}, in the ballpark of (\ref{ga}), have
been found in \cite{Narison,Prades,BERT98,Dortmund,Tajpei,PAPI99}.  The
result in \cite{BEL} corresponds to $\bsi=\bei=1$ (see section
\ref{sec:standard-model-parameters}) as the Dubna-DESY group has no
estimate of these non-perturbative parameters.  Recent reviews can be
found in \cite{CM00,BERT98,BUJA}.  Furthermore it has also been
suggested that final state interactions (FSI) could enhance $\epe$
by a factor of two \cite{BERT98,PAPI99,PA99}. A critical analysis of
this suggestion has been presented in \cite{AITAL}.  In our opinion
the issue of the size of FSI in the evaluation of $\epe$ is still an
open question. An interesting recent proposal in \cite{LL}, if
realized, could put the inclusion of FSI in the lattice calculations
of hadronic matrix elements in principle under control.

While all theoretical analyses quoted above use the NLO short distance
Wilson coefficients of \cite{NLOP,BJLW,ROMA1,AGH,BBL}, they differ in
the evaluation of the hadronic matrix elements of the relevant
operators.  Being of non-perturbative origin, the latter calculations
suffer from large theoretical uncertainties, which at present preclude
any firm prediction of $\epe$ within the SM and its extentions. In
this context, the matching between the short distance calculations of
Wilson coefficients and the long distance estimates of hadronic matrix
elements is not fully under control. In particular, it is crucial that
these two calculations are performed in the same renormalization
scheme \cite{AJBLAKE}. Moreover it has been pointed out recently in
\cite{donoghue00} that the presence of higher dimensional operators can
in the case of low matching scales complicate further this issue.
Finally, there is the question of isospin breaking effects that could
be more important than initially thought but are difficult to estimate
\cite{GV,ib}. There is a hope, however, that in the coming years the
situation may improve through advanced lattice simulations and new
analytical studies.

\begin{table}[tb]
\begin{center}
\begin{tabular}{||c|c||}\hline \hline
  {\bf Reference}&  $\epe~[10^{-4}]$ \\ \hline
Munich
\cite{EP99}&  $9.2^{+6.8}_{-4.0}$ (MC) \\
Munich
\cite{EP99}&   $1.4\to 32.7$ (S) \\
\hline
Rome
\cite{CM00,ROMA99}& $8.1^{+10.3}_{-9.5}$ (MC) \\
Rome
\cite{CM00,ROMA99}& $-13.0\to 37.0$ (S) \\
\hline
Trieste
\cite{BERT98}&  $22\pm 8 $ (MC) \\
Trieste
\cite{BERT98}&  $9\to 48 $ (S) \\
\hline
Dortmund
\cite{Dortmund}&    $6.8\to 63.9$ (S) \\
\hline
Montpellier
\cite{Narison}&  $ 24.2\pm 8.0$ \\
\hline
Granada-Lund
\cite{Prades}&   $ 34\pm 18$ \\
\hline
Dubna-DESY
\cite{BEL} 
&  $-3.2 \to 3.3$ (S) \\
\hline
Taipei
\cite{Tajpei}&  $7 \to 16$ \\
\hline
Barcelona-Valencia 
\cite{PAPI99}&  $17 \pm 6$ \\
\hline \hline
\end{tabular}
\caption[]{ Results for $\epe$ in the SM in  units of $10^{-4}$.
\label{tab:31738}}
\end{center}
\end{table}

In any case in view of the very large theoretical uncertainties and
sizable experimental errors there is still a lot of room for
non-standard contributions to $\epe$.  Indeed the results from NA31,
KTeV and NA48 prompted several analyses of $\epe$ within various
extensions of the SM like general supersymmetric models
\cite{Nierste,MM99,BS99,KN,SIL99}, models with anomalous gauge
couplings \cite{HE}, four-generation models \cite{Huang} and models
with additional fermions and gauge bosons \cite{Frampton}.
Unfortunately several of these extensions have many free parameters
and these studies are not very conclusive at present.  The situation
may change in the future when new data from high energy colliders will
restrict the possible ranges of parameters involved. For instance a
very recent analysis of the bounds on anomalous gauge couplings from
LEP2 data indicates that substantial enhancements of $\epe$ from this
sector are very unlikely \cite{TERR}.

On the other hand it appears that the $\epe$ data puts models 
in which there are new positive contributions to $\eps$ and 
negative contributions to 
$\varepsilon'$ in  difficulties. In particular 
as analyzed in \cite{EP99} the two Higgs Doublet Model II
can either be ruled out  with improved 
hadronic matrix elements or a  
powerful lower bound on $\tan\beta$ can 
be obtained from $\epe$.
In the Minimal Supersymmetric Standard Model (MSSM), in addition to charged
Higgs exchanges in loop diagrams, also charginos contribute.
For a suitable  choice of the
supersymmetric parameters, the chargino contribution
can enhance $\epe$  with respect to the SM expectations \cite{GG}.
Yet, as found in \cite{GG},
the most conspicuous effect of
minimal supersymmetry is a depletion of $\epe$.
The situation can be different in more general
models in which there are more parameters than
in the two Higgs doublet model II and in the MSSM, in particular
new CP violating phases.
As an example, in general supersymmetric models
$\epe$ can be considerably enhanced
through the contributions of the chromomagnetic
penguins \cite{Nierste,MM99,BS99,GMS}, $Z^0$-penguins
with the opposite sign to the one in the SM
\cite{BS99,ISI,BS98,CHANOWITZ} and isospin breaking effects \cite{KN}.

The purpose of the present paper is a new analysis within the MSSM
with minimal flavour violation,
which updates and generalizes the work of Gabrielli and
Giudice \cite{GG}. Compared to the latter paper, the main
new ingredients of our analysis are:

\begin{itemize}
\item
The inclusion of NLO-QCD and QED corrections;
\item
the update in the CKM and non-perturbative parameters;
\item the imposition of constraints on supersymmetric parameters
  coming in particular from $\Delta\varrho$ in the electroweak
  precision studies and from the lower bound on the neutral Higgs mass,
  in addition to the ones provided by the $B\to X_s\gamma$ decay rate;
\item
the simultaneous study of the rare decays
$\kpnn$, $\klpn$, $\klee$, $\bnunu$ and $\bmumu$ which are sensitive
to $Z^0$-penguin contributions.
\end{itemize}

As in \cite{GG} we will consider the MSSM with minimal flavour
violation, in which the effective local operators are the same as in
the SM and the only source of CP violation is the KM phase. Since the
supersymmetric particles have masses ${\cal O}(\mw)$ or higher and the
effective low energy operators are the same as in the SM, the impact
of new contributions in a renormalization group analysis is only felt
in the initial conditions for the Wilson coefficients taken usually at
$\mu={\cal O}(\mw)$.  The renormalization group transformation from
$\mu={\cal O}(\mw)$ down to $\mu={\cal O}(1~\gev)$ is on the other
hand the same as in the SM. This simplifies the inclusion of QCD
corrections considerably.

In view of the fact that the new contributions enter only the initial 
conditions for the Wilson coefficients it is useful to follow
\cite{PBE} and to cast the expression for $\epe$ into a formula
which exhibits transparently the new supersymmetric contributions.
One finds \cite{EP99} 
\begin{equation}        \label{PE}
        \epe=\IM\lambda_t \cdot F_{\varepsilon'}\,,
                \qquad \qquad F_{\varepsilon'}= P_0+\sum_r P_r F_r
\end{equation}
where $\lambda_t=V_{td}V_{ts}^*$.  The coefficients $P_i$ depend on
non-perturbative parameters and include NLO-QCD and QED corrections.
They are common to the SM and the MSSM.  The functions $F_r$ resulting
from various box and penguin diagrams contain both the SM and
supersymmetric contributions involving new exchanges such as charged
Higgs particles, charginos, squarks etc.  They depend, in addition to
$\mt$, on the masses of the new particles as well as on a number of
other supersymmetric parameters. The explicit expressions for $F_r$
and the numerical values for the coefficients $P_i$ will be given in
sections \ref{sec:basic-functions} and \ref{sec:compendium}
respectively.

There are many new contributions in the MSSM
such as charged Higgs, chargino, neutralino and gluino contributions.
However, in the case of
minimal flavour and CP violation it is a good approximation to keep only
charged Higgs and chargino contributions.

Our paper is organized as follows. In section \ref{sec:mssm} we review
the elements of the MSSM which are relevant for our analysis. In
section \ref{sec:basic-functions} we give the list of all $F_r$
functions which can be 
decomposed into W, charged-Higgs and chargino contributions
\begin{equation}
        F_r=(F_r)_{SM}+(F_r)_H+(F_r)_\chi\equiv (F_r)_{SM}+(F_r)_{SUSY}\,,
\end{equation}
and we compare our results with those given in the literature.  In
section \ref{sec:compendium} we collect the formulae for
$\varepsilon$, $B^0_{d,s}-\bar B^0_{d,s}$ mixings, $\epe$ and the rare
decays in question.  In section \ref{sec:general-considerations} some
general considerations are given.  In particular we address the issue
of a {\it universal unitarity triangle} \cite{UUT} for all models,
like the SM and the constrained MSSM, that do not have any new phases
beyond the CKM phase and no new operators.  In section
\ref{sec:constraints} we discuss constraints on supersymmetric
parameters coming from the $B\to X_s\gamma$ decay, $\Delta\varrho$ in
the electroweak precision studies and the lower bound on the neutral
Higgs mass.  The numerical analysis of the formulae of section
\ref{sec:compendium} including the constraints from section
\ref{sec:constraints} is presented in section
\ref{sec:numerical-analysis}.  Finally section \ref{sec:summary}
contains a summary and conclusions. A list of auxiliary functions is
given in the Appendix.

\section{The Minimal Supersymmetric Standard Model}
\label{sec:mssm}
\setcounter{equation}{0}
\subsection{FCNC Processes and CP Violation in the MSSM}

In general, the MSSM contains more than a hundred new parameters with
respect to the SM. A large number of these parameters can be a new
source of Flavour-Changing Neutral Currents (FCNC) and of CP
violation.  Phenomenological studies
\cite{GMS,BERTOL,Dugan:1985qf,Hall:1986dx,Gabbiani:1989rb,Hagelin:1994tc}
have shown that most of these new flavour- and CP-violating parameters
are constrained to be very small by the experimental results on FCNC
and CP violation. One is then left with the so-called ``flavour
problem'' and ``CP problem'', i.e. with the difficulty of explaining
the smallness of these new parameters. We do not discuss here the
possible solutions to these problems, but rather assume that they have
been solved leading to a version of the MSSM where no new flavour- and
CP-violating parameters appear (minimal flavour violation). Although
this is not a necessary outcome of the solution of the flavour and CP
problems, it is certainly a well-defined and interesting scenario that
leads to ``minimal'' modifications of the SM predictions on FCNC and
CP violating processes. 

\subsection{The Model}
\label{subsec:the-model}

Let us now specify the details of the MSSM with minimal flavour and CP
violation.
After applying to the superfields those field redefinitions that
diagonalize the quark mass matrices $M_u$ and $M_d$, the squark mass
matrices read
\begin{equation}                \label{squarkmassu}
        \tM^2_u = \left(
                \begin{array}{cc}
                        \hspace{0.5cm}
                        \tM^2_{\sq_L} + \left( M_u \right)^2
                                +M^2_Z \cos 2 \beta \,\,\alpha_L^{(u)}
                                \hspace{0.5cm} &
                        \left( -\mu^*  \cot \beta + A_u^\dagger \right)
                                M_u                                     \\
                        M_u \left(- \mu  \cot \beta +  A_u \right) &
                        \tM^2_{\supq_R} + \left(M_u\right)^2
                                + M^2_Z \cos 2 \beta \,\,\alpha_R^{(u)}
                \end{array}
        \right)
\end{equation}
\begin{equation}                        \label{squarkmassd}
        \tM^2_d = \left(
                \begin{array}{cc}
                        V^\dagger \tM^2_{\sq_L} V
                                + \left( M_d \right)^2
                                + M^2_Z \cos 2 \beta \,\,\alpha_L^{(d)}
                                \hspace{0.3cm} &
                        \left( -\mu^* \tan\beta + A_d^\dagger \right)
                                M_d                     \\
                        M_d \left( -\mu \tan \beta +  A_d \right) &
                        \tM^2_{\sdown_R} + \left( M_d \right)^2
                                + M^2_Z \cos 2 \beta \,\,\alpha_R^{(d)}
                \end{array}
        \right)
\end{equation}
where each entry denotes a three-by-three submatrix and where
\begin{equation}
        \begin{array}{lcl}
                \alpha_L^{(u)} = \half - \frac{2}{3} \sin^2 \Theta_{\rm W} &
                \hspace{2cm} &
                \alpha_R^{(u)} = \frac{2}{3} \sin^2 \Theta_{\rm W} \\
                \alpha_L^{(d)} = -\half + \frac{1}{3} \sin^2 \Theta_{\rm W} &
                \hspace{2cm} &
                \alpha_R^{(d)} = -\frac{1}{3} \sin^2 \Theta_{\rm W}\,.
        \end{array}
\end{equation}
The fact that the explicit squark mass matrices in the two left-left
sectors are related by the CKM matrix $V$ is a remnant of $SU(2)$
gauge symmetry.
The \mbox{$3 \times 3$} matrices
$\tM^2_{\sq_L}$, $\tM^2_{\supq_R}$, $\tM^2_{\sdown_R}$, $A_u$, $A_d$ are
a priori unconstrained complex matrices (except for the hermiticity of
$\tM^2_{\sq_L}$, $\tM^2_{\supq_R}$, $\tM^2_{\sdown_R}$) parameterizing the
soft supersymmetry breaking in the squark sector. We also define,
as usual, $\tan \beta = v_2/v_1$, while $\mu$ denotes the supersymmetric
Higgs mixing parameter.

In the basis chosen, all neutral-current couplings (couplings of neutral
gauge bosons, neutral scalars, and neutral higgsinos and gauginos
to quarks and squarks) are still diagonal
in the sense that they do not mix members of different quark supermultiplets,
while the charged current couplings contain the CKM
matrix in the same way as in the SM.
However, FCNC arise after diagonalization of the squark mass matrices as soon
as any one of the eight \mbox{$3 \times 3$} submatrices of
eqs.~(\ref{squarkmassu}) and~(\ref{squarkmassd})
is nondiagonal, while new CP violation effects are present 
if they are not real.
Furthermore, we can only have $\tM^2_{\sq_L}$ and $V^\dagger \tM^2_{\sq_L} V$
diagonal at the same time if
\begin{equation}
        \tM^2_{\sq_L} = \tm^2 {\bf 1}\,.
\label{eq:univleft}
\end{equation}
Our assumption of minimal flavour and CP violation is consequently
equivalent to real $\mu$, real and diagonal $A_u$, $A_d$,
$\tM^2_{\supq_R}$, $\tM^2_{\sdown_R}$, and $\tM^2_{\sq_L} = \tm^2 {\bf
  1}$. 

The elements of $A_u$, $A_d$ except the one multiplying the top quark mass
are further constrained to be very small
by requiring that the model does not develop charge- or color-breaking VEVs
or potentials unbounded from below \cite{CCB1,CCB2,CCB3}, 
so that we may neglect
them for all squarks but the stops.
Similarly, if $\tan \beta$ is not
too large, the off-diagonal terms
proportional to $\mu$ are negligible for $\sq \not= \stp$.
In this case, the squark mass matrices are
already diagonal except for mixing in the stop sector.
The real, symmetric stop
mass matrix is diagonalized via the orthogonal transformation
\beq
T \tilde{M}^2_t T^T = \left( \begin{array}{cc}
                \tm^2_{t_1} & 0 \\
                0 & \tm^2_{t_2}
        \end{array} \right) ,~~~
T =  \left( \matrix{
        \cos \tht & \sin \tht \cr
        -\sin \tht & \cos \tht \cr}
\right)~.
\eeq

As a further simplifying assumption, we will set all other squark masses
equal, which then implies $\tm^2_{i_L,i_R} \approx \tm^2$ for $i=u,d,s,c,b$.
A strict equality is inconsistent with~(\ref{squarkmassu}) and
(\ref{squarkmassd})
for the ``left-handed'' squarks because of the different D-terms 
in the up and down sectors, but this difference is
numerically insignificant  and will be ignored here.
We therefore set
\begin{equation}                \label{degenerate_squarks}
        \tm^2_{i_L,i_R} = \tm^2, \hspace{2cm} i=u,d,s,c,b.
\end{equation}
We then take the input parameters in the squark sector
to be the two stop masses and the stop mixing angle $\tht$. From the above
we also have
\begin{equation}                \label{squark_stop_relation}
        \tm^2 = -m_t^2 + \tm_{t_1}^2 \cos^2 \tht + \tm_{t_2}^2 \sin^2 \tht
\end{equation}
as a formula for computing $\tm$. We choose $\tilde t_2$ to be
the light stop.

In the leptonic sector we neglect Left-Right mixing, and assume,
analogously to (\ref{eq:univleft}) and  (\ref{degenerate_squarks}),
\begin{equation}
  \label{eq:degenerate_sleptons}
        \tm^2_{i_L} = \tm_\ell^2, \hspace{2cm} i=\nu,e.  
\end{equation}

The chargino mass matrix is diagonalized by two orthogonal
$2\times 2$ matrices
$\tilde U$ and $\tilde V$, according to:
\beq\label{UV}
\tilde U \left(
\matrix{
M & m_{\W}\sqrt{2}\sin{\beta} \cr
m_{\W}\sqrt{2}\cos{\beta} & \mu \cr}
\right) {\tilde V}^T=
\left(
\matrix{
\tilde{m}_{\chi_1} & 0 \cr
0 & \tilde{m}_{\chi_2} \cr}
\right) ,
\nonumber
\eeq
where $\mu$ and $\tan{\beta}$ have been introduced before and $M$ is the
weak gaugino mass. $M$ and $\mu$
are taken to be real to avoid the appearance of new sources of CP
violation.
We prefer to choose the lightest chargino mass, which we call
$m_{\chi_1}$, as an input parameter instead of $M$.

In the Higgs sector, the only additional relevant parameter is the mass of the
charged Higgs, $m_{H^+}$.

Altogether, we are left with the seven input parameters $m_{\stp_1}$,
$m_{\stp_2}$, $\tht$, $m_{\chi_1}$, $\mu$, $\tan{\beta}$, and
$m_{H^+}$.  The ranges for these parameters used in our numerical
analysis are discussed in section \ref{sec:susy-parameter-space}. In
the computation of the rare decays $\kpnn$, $\klpn$, $\klee$, $\bnunu$
and $\bmumu$ one more input is needed, the slepton mass $m_{\tilde
  \ell}$.

\subsection{Chargino Interactions}
Let us denote by $d$ the quarks $(d,s,b)$ and by $\tilde u_k$ the
squarks $(\tilde u,\tilde c,\tilde t)$ with $k=1,2$. The interaction of
charginos $(\tilde\chi^+_j,~j=1,2)$ with $d$ and $\tilde u_k$ is then given
explicitly as follows:
\begin{equation}\label{LX}
{\cal L}_\chi = g V^*_{ud} 
\bar d \left(Z_{jk}P_L+Y_{jk}P_R\right) {\rm C}
        \left(\overline{\tilde \chi^{+}_j} \right)^T \tilde u_k
+{\mathrm h.c.}
\end{equation}
where $P_{L,R}=(1\mp\gamma_5)/2$, while $\rm C$ is the charge conjugation
matrix. The couplings $Z_{jk}$ and $Y_{jk}$ are
real so that the only phases are present in the CKM elements. We
have:
\begin{equation}
Y_{j1}=\frac{m_u}{\sqrt{2}M_W\sin\beta}\tilde V_{j2}\sin\theta_{\tilde u}
        -\tilde V_{j1}\cos\theta_{\tilde u}
\end{equation}
\begin{equation}
Y_{j2}=\frac{m_u}{\sqrt{2}M_W\sin\beta}\tilde V_{j2}\cos\theta_{\tilde u}
        +\tilde V_{j1}\sin\theta_{\tilde u}
\end{equation}
\begin{equation}
Z_{j1}=\frac{m_d}{\sqrt{2}M_W\cos\beta}\tilde U_{j2}\cos\theta_{\tilde u}
        = 0
\end{equation}
\begin{equation}
Z_{j2}=-\frac{m_d}{\sqrt{2}M_W\cos\beta}\tilde U_{j2}\sin\theta_{\tilde u}
        = 0
\end{equation}
with $\tilde V_{ji}$ and $\tilde U_{ji}$  defined in (\ref{UV}).
The $m_u$ denotes generally the mass of the up-quark $(u,c,t)$
corresponding to a given up-squark $\tilde u$. 
$\theta_{\tilde u}$ are the mixing angles in the squark sector.
Note that since we neglect quark masses other than that of the top,
 we have $\theta_{\tilde u}=0$ 
for $\tilde u\not=\tilde t$ (see the preceding
section) as well as $Z_{jk} = 0$. An exception is the $B\to X_s\gamma$
decay in which $m_b$ and consequently also $Z_{jk}$ cannot be set to zero.

The Lagrangian (\ref{LX}) applies also to lepton-sneutrino-chargino
interactions after the CKM factor has been removed and $\theta_{\tilde u}$,
$Y_{j2}$ and $Z_{jk}$ all set to zero. 
If the masses of charged leptons are neglected only the couplings
$Y_{j1}=-\tilde V_{j1}$ are non-vanishing in this case.

Similarly the interaction of charginos with $u$ and $\tilde d_k$ is
given by
\begin{equation}\label{LXU}
{\cal L}_\chi = g V_{ud} 
\bar u 
\left(\bar Z_{jk}P_L+\bar Y_{jk}P_R\right)\tilde\chi^{+}_j\tilde d_k
+{\mathrm h.c.}
\end{equation}
where
\begin{equation}
\bar Y_{j1}= -\tilde U_{j1}
\end{equation}
\begin{equation}
        \bar Y_{j2} = \bar Z_{j1} = \bar Z_{j2} = 0\,.
\end{equation}

The Lagrangian (\ref{LXU}) describes, mutatis mutandis, the
neutrino-slepton-chargino interactions once the CKM factor has been removed.

\section{Basic Functions}
\label{sec:basic-functions}
We list next the basic functions which enter the formulae for
$\varepsilon$, $B^0-\bar B^0$ mixing, $\epe$ and rare decays
in section \ref{sec:compendium}.
The functions relevant for $\varepsilon$ and 
$B^0-\bar B^0$ mixing are given
as follows:
\bea
S_{SM}(c,t)&=&S(x_{c\W},x_{t\W})\nonumber \\
S_{\HH}(c,t)&=&\frac{x_{\HH\W}}{4\tan^4{\beta}}
L_2(x_{c\HH},x_{t\HH},1)+
\frac{2}{\tan^2{\beta}}\left[\frac{1}{4}L_2(x_{c\W},
x_{t\W},x_{\HH\W})-
L_1(x_{c\W},x_{t\W},x_{\HH\W})\right]\nonumber \\
S_{\chi}(t,t)&=&f(t,t)+f(c,c)-2 f(t,c)\nonumber \\
f(c,t)&=&
\sum_{\stackrel{i,j=1,2}{\scriptscriptstyle{h,k=1,2}}}
x_{\W\chi_j}
Y_{i\tilde{c}_h}Y_{i\tilde{t}_k}
Y_{j\tilde{c}_h}Y_{j\tilde{t}_k}
L_3(x_{\tilde{t}_{k}\chi_j},x_{\tilde{c}_{h}\chi_j},x_{\chi_i\chi_j}) 
\eea
where
$x_{ab}\equiv m_a^2/m_b^2$.
The functions $S$ and $L_{1,2,3}$ are given in the  appendix.
These results agree with \cite{BERTOL,BBHLS,BRANCO}. They agree also
with \cite{GG} after the erratum has been taken into
account. 

The functions which include the contributions from photon-penguins ($D$),
Z-penguins ($C$),  boxes with external $d\bar d$
($B^{(d)}$), $u\bar u$ ($B^{(u)}$), $\nu\bar\nu$ ($B^{(\nu\bar \nu)}$),
 $e\bar e$ ($B^{(e\bar e)}$)  and 
gluon-penguins ($E$), are given as follows:
%
%
\beq
D=D_{SM}(x_{t\W})+\frac{1}{\tan^2\beta}D_H(x_{t\HH})+
2 \sum_{\stackrel{j=1,2}{\scriptscriptstyle{k=1,2}}}
\left[
Y^2_{j\tilde{t}_k}x_{\W\tilde{t}_k}D_{\chi}(x_{\chi_j \tilde{t}_k})-
\left(\tilde{t}\rightarrow \tilde{c}\right)\right]
\label{fping}
\eeq
%
%
\bea
C&=&C_{SM}(x_{t\W})+\frac{x_{t\W}}{\tan^2\beta}C_H(x_{t\HH})+
2 \sum_{\stackrel{i,j=1,2}{\scriptscriptstyle{h,k=1,2}}}
\left[
Y_{j\tilde{t}_{k}}Y_{i\tilde{t}_{h}}
\left\{
  \delta_{ij}\Delta^{\tilde t}_{hk}
C_{\chi}^{(1)}(x_{\tilde{t}_h\chi_j},x_{\tilde{t}_k\chi_j})
\right.\right.
\nonumber \\*
&+&
\left.\left.
\delta_{hk}\left[
\tU_{i1} \tU_{j1}C^{(2)}_{\chi}(x_{\chi_j\tilde{t}_k},x_{\chi_i\tilde{t}_k})-
\tV_{i1} \tV_{j1}
C^{(1)}_{\chi}(x_{\chi_j\tilde{t}_k},x_{\chi_i\tilde{t}_k})
\right]\right\}-\left(\tilde{t}\rightarrow\tilde{c}\right)\right]
\nonumber \\*
&+&
\frac{1}{8}\left[\log x_{\tilde{t}_1\tilde{c}_1}
+\sin^2\theta_{\tilde{t}}\log  x_{\tilde{t}_2\tilde{t}_1}\right]
\label{zping}
\eea
where 
\begin{equation}
\Delta^{\tilde t}_{hk}=\left\{ \begin{array}{ll}
\cos^2 \theta_{\tilde{t}} & k=h=1 \\
-\sin\theta_{\tilde{t}}\cos\theta_{\tilde{t}} & k\not=h\\
\sin^2 \theta_{\tilde{t}}& k=h=2 \end{array} \right.
\end{equation}
%
%
\bea
B^{(d)}&=&B_{SM}(x_{t\W})+
\frac{1}{4}\sum_{\stackrel{i,j=1,2}{\scriptscriptstyle{k=1,2}}}
{\tV}_{j1} {\tV}_{i1}
x_{\W\chi_j}\left[
Y_{j\tilde{t}_{k}}Y_{i\tilde{t}_{k}}
B_{\chi}^{(d)}(x_{\tilde{t}_k\chi_j},x_{\tilde{u}_1\chi_j},x_{\chi_i\chi_j})
\right.\nonumber \\*
& & - \left.
\left(\tilde{t}\rightarrow\tilde{c}\right)\right]
\label{dbox}    \\
%
%
B^{(u)}&=&B_{SM}(x_{t\W})+
\frac{1}{8}\sum_{\stackrel{i,j=1,2}{\scriptscriptstyle{k=1,2}}}
\tU_{j1} \tU_{i1}x_{\W\chi_j}\sqrt{x_{\chi_i\chi_j}}\left[
Y_{j\tilde{t}_{k}}Y_{i\tilde{t}_{k}}
B_{\chi}^{(u)}(x_{\tilde{t}_k\chi_j},x_{\tilde{d}_1\chi_j},x_{\chi_i\chi_j})
\right. \nonumber \\*
& & - \left.
\left(\tilde{t}\rightarrow\tilde{c}\right)\right]
\label{ubox}    \\
%
%
B^{(\nu\bar\nu)}&=&B_{SM}(x_{t\W})+
\frac{1}{8}\sum_{\stackrel{i,j=1,2}{\scriptstyle{k=1,2}}}
\tU_{j1} \tU_{i1}x_{\W\chi_j}\sqrt{x_{\chi_i\chi_j}}\left[
Y_{j\tilde{t}_{k}}Y_{i\tilde{t}_{k}}
B_{\chi}^{(u)}(x_{\tilde{t}_k\chi_j},x_{\tilde{e}_1\chi_j},x_{\chi_i\chi_j})
\right. \nonumber \\*
& & - \left.
\left(\tilde{t}\rightarrow\tilde{c}\right)\right]
\label{nubox}   \\
%
%
B^{(e\bar e)}&=&B_{SM}(x_{t\W})+
\frac{1}{4}\sum_{\stackrel{i,j=1,2}{\scriptstyle{k=1,2}}}
\tV_{j1} \tV_{i1}x_{\W\chi_j}\left[
Y_{j\tilde{t}_{k}}Y_{i\tilde{t}_{k}}
B_{\chi}^{(d)}(x_{\tilde{t}_k\chi_j},x_{\tilde{\nu}\chi_j},x_{\chi_i\chi_j})
\right.\nonumber \\*
& & - \left.
\left(\tilde{t}\rightarrow\tilde{c}\right)\right]
\label{mubox}   \\
%
%
E &=& E_{SM}(x_{t\W})+\frac{1}{\tan^2\beta}E_{H}(x_{t\HH})+
2\sum_{\stackrel{j=1,2}{\scriptscriptstyle{k=1,2}}}
\left[Y^2_{j\tilde{t}_{k}}x_{\W\tilde{t}_k}
E_{\chi}(x_{\chi_j\tilde{t}_k})-\left(\tilde{t}\rightarrow\tilde{c}\right)
\right]\,.
\label{gpeng}
\eea

The results for the SM and charged Higgs contributions
are the same as in \cite{BBHLS} and agree with most papers in the literature.
The chargino contribution to the $D$ function given here agrees with 
\cite{BERTOL,MISIAK} and also with the erratum
in \cite{GG}. Our result for the chargino
contribution to the $C$ function
agrees with~\cite{BERTOL,MISIAK} and the erratum in~\cite{GG}. 
The $C$ functions given in these  papers at first sight seem to differ
from each other and from our result~(\ref{zping}).
However, after the summations over $i$,$j$,$h$,$k$ have been performed and
the $\tilde c$ part subtracted, the results of these papers reduce
to our result. In particular the constant terms in~\cite{MISIAK} disappear.
Note that the convention for
$\theta_{\tilde t}$ in~\cite{MISIAK} differs from ours by a sign.

Next our chargino contribution to  $B^{(d)}$ differs
from the result of~\cite{GG} by an overall sign, while we find
the chargino contribution to $B^{(u)}$ to be a factor of 2 larger than
given in the latter paper. 
Finally our result for the chargino contribution to  $B^{(e\bar e)}$ agrees
with~\cite{MISIAK} and differs by an overall sign from~\cite{BERTOL}. 
In view of the fact that these box contributions are small these
differences are unimportant for the whole analysis.

\section{Compendium of Phenomenological Formulae}
\label{sec:compendium}
\setcounter{equation}{0}
\subsection{Basic Formula for $\eps$}
            \label{subsec:epsformula}
The indirect CP violation in $K \to \pi\pi$ is described by the well
known parameter $\eps$. The general formula for $\eps$ is given as
follows
\begin{equation}
\eps = \frac{\exp(i \pi/4)}{\sqrt{2} \Delta M_K} \,
\left( \IM M_{12} + 2 \xi \RE M_{12} \right)
\label{eq:epsdef}
\end{equation}
where
\begin{equation}
\xi = \frac{\IM A_0}{\RE A_0}
\label{eq:xi}
\end{equation}
with $A_0 \equiv A(K \to (\pi\pi)_{I=0})$ and $\Delta M_K$ being
the $K_L$-$K_S$ mass difference. The off-diagonal element $M_{12}$ in
the neutral $K$-meson mass matrix represents the $K^0$-$\bar K^0$
mixing. It is given in the MSSM by
\begin{equation}
M_{12} = \frac{G_F^2}{12 \pi^2} F_K^2 \hat B_K m_K M_W^2
\left[ {\lambda_c^*}^2 \eta_1 F_{cc} + {\lambda_t^*}^2 \eta_2 F_{tt} +
2 {\lambda_c^*} {\lambda_t^*} \eta_3 F_{ct} \right]
\label{eq:M12K}
\end{equation}
where $\lambda_i=V_{is}^*V_{id}$ and 
\begin{equation}
F_{cc}=S_{SM}(c,c)=x_{cW}
\end{equation}
\begin{equation}\label{tt}
F_{tt}=S_{SM}(t,t)+S_H(t,t)+S_{\chi}(t,t)
\end{equation}
\begin{equation}
F_{ct}=S_{SM}(c,t)+\frac{\eta^H_3}{\eta_3} S_H(c,t)~.
\end{equation}
In what follows we neglect the charged Higgs contribution to $F_{cc}$.
Due to the assumed mass degeneracy for squarks in the first two generations,
chargino exchanges contribute only to $F_{tt}$. In what follows we
will use the NLO results of \cite{NLOS1,NLOS2}
for the QCD factors $\eta_i$:
\begin{equation}
\eta_1 = 1.38 
\qquad
\eta_2 = 0.57
\qquad
\eta_3 = 0.47 \, .
\label{eq:etaknum}
\end{equation}
as obtained in the SM. It should be stressed that in the leading
logarithmic approximation (LO) $\eta_2$ is indeed the same for all three
contributions in (\ref{tt}). At NLO however this is no longer true
because then also gluon corrections to various box diagrams have to be
included and these are different for the three contributions in (\ref{tt}).
These corrections are unknown at present.
Since in the SM the NLO corrections to $\eta_2$ mainly
remove the renormalization scale ambiguities present in NLO, we think
that in view of other uncertainties in the supersymmetry sector this
procedure of using $\eta_2$ at NLO is justified. In the case of $F_{ct}$
the QCD corrections to charged Higgs exchanges differ from the SM 
results already in the leading order and are given by  $\eta^H_3=0.21$
\cite{BBHLS}.

Finally
$\hat B_K$ is a well known non-perturbative parameter,
$F_K$ is the $K$-meson decay constant and $m_K$
the $K$-meson mass. 
The last term in (\ref{eq:epsdef}) constitutes at
most a 2\,\% correction to $\eps$ and consequently can be neglected
in view of other uncertainties, in particular those connected with
$\hat B_K$.  

\subsection{Basic Formula for $B^0-\bar B^0$ Mixing}
            \label{subsec:BBformula}
The $B^0-\bar B^0$ mixing is usually described by
$(\Delta M)_{d,s}$, the mass difference between the mass
eigenstates in the $B_d^0-\bar B_d^0$ system and the $B_s^0-\bar B_s^0$
system, respectively.
$(\Delta M)_{d,s}$ is given as follows:
\begin{equation}
(\Delta M)_{d,s} = \frac{G_F^2}{6 \pi^2} \eta_B m_{B_{d,s}} 
(\hat B_{B_{d,s}} F_{B_{d,s}}^2 ) M_W^2 F_{tt} |V_{t(d,s)}|^2
\label{eq:xds}
\end{equation}
where $F_{tt}$ is again the function given in (\ref{tt}).
$\hat B_B$ is a non-perturbative parameter analogous to $\hat B_K$, $F_B$ is
the B-decay constant and 
$\eta_B$ the QCD factor given at NLO in the SM by $\eta_B=0.55$ 
\cite{NLOS1,NLOS3}. Supersymmetric corrections modify this factor
first at the NLO level. As they are unknown at present we will
use the SM value for $\eta_B$ also in the MSSM.
\subsection{Basic Formula for $\epe$}
Using the functions listed in Section \ref{sec:basic-functions}, the
formula for $\epe$ of \cite{EP99} generalizes to the supersymmetric
case as follows: \be \frac{\varepsilon'}{\varepsilon}= \IM\lambda_t
\cdot F_{\varepsilon'}
\label{epeth}
\ee
where
\be
F_{\varepsilon'} =P_0 + P_X \, X + P_Y \, Y + P_Z \, Z+ P_E \, E
\label{FE}
\ee
with
\begin{equation}
X=C-4 B^{(u)},\qquad Y=C-B^{(d)},\qquad Z=C+\frac{1}{4}D.
 \label{eq:3b}
\end{equation}
The functions $D$, $C$, $B^{(d)}$, $B^{(u)}$ and $E$ are listed
in (\ref{fping})--(\ref{gpeng}). Retaining only the SM contributions
in the latter functions one obtains the formula (2.38) of \cite{EP99}.

The coefficients $P_i$ are given in terms of the non-perturbative
parameters $B_6^{(1/2)}$ and $B_8^{(3/2)}$ and the strange quark
mass  $\ms(\mc)$ as follows:
\begin{equation}
P_i = r_i^{(0)} + 
r_i^{(6)} R_6 + r_i^{(8)} R_8 \, 
\label{eq:pbePi}
\end{equation}
where
\be\label{RS}
R_6\equiv \bsi\left[ \frac{137\mev}{\ms(\mc)+\md(\mc)} \right]^2,
\qquad
R_8\equiv \bei\left[ \frac{137\mev}{\ms(\mc)+\md(\mc)} \right]^2.
\ee
$\bsi$ and $\bei$ parameterize the matrix elements of the dominant
QCD-penguin ($Q_6$) and the dominant electroweak penguin ($Q_8$)
operator respectively.
The numerical values of $r_i^{(0)}$, $r_i^{(6)}$ and $r_i^{(8)}$ 
for different values of $\Lms^{(4)}$ at $\mu=\mc$ in the NDR 
renormalization scheme are given in table~\ref{tab:pbendr}. 
This table differs from the one presented in \cite{EP99}
in that $\OEE=0.25$ has been replaced by $\OEE=0.16$ in accordance
with the most recent calculation in \cite{ECKER99} which
gives $\OEE=0.16\pm0.03$ (see however \cite{GV}). All other input
parameters are as in \cite{EP99} 
where further details connected with the coefficients in 
 (\ref{eq:pbePi}) and the parameters $\bsi$ and $\bei$ can be found.

\begin{table}[tb]
\begin{center}
\begin{tabular}{|c||c|c|c||c|c|c||c|c|c|}
\hline
& \multicolumn{3}{c||}{$\Lms^{(4)}=290\mev$} &
  \multicolumn{3}{c||}{$\Lms^{(4)}=340\mev$} &
  \multicolumn{3}{c| }{$\Lms^{(4)}=390\mev$} \\
\hline
$i$ & $r_i^{(0)}$ & $r_i^{(6)}$ & $r_i^{(8)}$ &
      $r_i^{(0)}$ & $r_i^{(6)}$ & $r_i^{(8)}$ &
      $r_i^{(0)}$ & $r_i^{(6)}$ & $r_i^{(8)}$ \\
\hline
0 &
  --3.122  &   10.905  &    1.423  &
  --3.167  &   12.409  &    1.262  & 
  --3.210  &   14.152  &    1.076  \\
$X$ &
    0.556  &    0.019  &    0      &  
    0.540  &    0.023  &    0      &  
    0.526  &    0.027  &    0      \\
$Y$ &
    0.404  &    0.080  &    0      & 
    0.387  &    0.088  &    0      & 
    0.371  &    0.097  &    0      \\
$Z$ &
    0.412  &  --0.015  &  --9.363  & 
    0.474  &  --0.017  & --10.186  & 
    0.542  &  --0.019  & --11.115  \\
$E$ &
    0.204  &  --1.276  &    0.409  & 
    0.188  &  --1.399  &    0.459  &  
    0.172  &  --1.533  &    0.515  \\
\hline
0 &
  --3.097  &    9.586  &    1.045  &
  --3.141  &   10.748  &    0.867  &
  --3.183  &   12.058  &    0.666  \\
\hline
\end{tabular}
\end{center}
\caption[]{Coefficients in the formula (\ref{eq:pbePi})
 for various $\Lms^{(4)}$ in 
the NDR scheme.
The last row gives the $r_0$ coefficients in the HV scheme.
\label{tab:pbendr}}
\end{table}

\subsection{ Rare K and B Decays}

The well known expressions for rare K and B decays in the SM
can easily be generalized to the corresponding expressions
in the MSSM. We have \cite{AJBLAKE}

\begin{equation}\label{bkpn}
Br(\kpnn)=1.54\cdot 10^{-4}
\cdot\left[\left(\imlt\cdot X^{(\nu\bar\nu)}\right)^2+
\left(\relc\cdot\bar X_c+\relt\cdot X^{(\nu\bar\nu)}\right)^2
\right]~,
\end{equation}
where  $\bar X_c=(9.8\pm1.4)\cdot 10^{-4}$ 
stands for the internal charm 
contribution evaluated at NLO in the SM \cite{BB9499}. 
$\relc=-\lambda(1-\lambda^2/2)$ with $\lambda=0.22$.
The supersymmetric contributions reside only in $X^{(\nu\bar\nu)}$.

\begin{equation}\label{bklpn}
Br(K_{\rm L}\to\pi^0\nu\bar\nu)=6.8\cdot 10^{-4}\cdot
\left(\imlt\cdot X^{(\nu\bar\nu)}\right)^2
\end{equation}

\begin{equation}\label{9a}
Br(K_{\rm L} \to \pi^0 e^+ e^-)_{\rm dir} = 6.3\cdot 10^{-6}(\IM\lambda_t)^2
(\tilde y_{7A}^2 + \tilde y_{7V}^2)\,,
\end{equation}
where
\begin{equation}\label{y7vpbe}
\tilde{y}_{7V} =
P_0 + \frac{Y^{(e\bar e)}}{\sin^2\Theta_{\rm W}} - 4 Z~,\qquad
\tilde{y}_{7A}=-\frac{1}{\sin^2\Theta_{\rm W}} Y^{(e\bar e)}
\end{equation}
with $P_0=3.05\pm 0.08$ including NLO corrections \cite{BLMM}
and $\sin^2\Theta_{\rm W}=0.23$. In (\ref{9a}) we have only
given the directly CP violating contribution to this decay.

Finally
\begin{equation}
Br(B \to X_s \nu\bar\nu) = 1.5 \cdot 10^{-5} \,
\frac{|V_{ts}|^2}{|V_{cb}|^2} \,
\left(X^{(\nu\bar\nu)}\right)^2 \, ,
\label{eq:bxsnnnum}
\end{equation}

\begin{equation}\label{bbmmnum}
Br(B_s\to\mu^+\mu^-)=3.3\cdot 10^{-9}\left[\frac{\tau(B_s)}{1.6
\mbox{ps}}\right]
\left[\frac{F_{B_s}}{210\mev}\right]^2 
\left[\frac{|V_{ts}|}{0.040}\right]^2 
\left( Y^{(e\bar e)} \right)^2
\end{equation}
where $\tau(B_s)$ and $F_{B_s}$ is the life-time and the decay constant
of the $B_s$ meson respectively.

The functions $ X^{(\nu\bar\nu)}$ and $Y^{(e\bar e)}$ are given as follows
\begin{equation}
X^{(\nu\bar\nu)}=C-4 B^{(\nu\bar\nu)},\qquad 
Y^{(e\bar e)}=C-B^{(e\bar e)}.
 \label{eq:XYSUSY}
\end{equation}
with $C$, $B^{(\nu\bar\nu)}$, $B^{(e\bar e)}$ listed in 
(\ref{zping}), (\ref{nubox}) and
(\ref{mubox}) respectively. The function $Z$ is defined in (\ref{eq:3b}).

\section{General Considerations}
\label{sec:general-considerations}
\setcounter{equation}{0}
\subsection{Strategy A: Universal Unitarity Triangle}
The SM, the MSSM and any model in which all flavour changing
transitions are governed by the CKM matrix with no new phases beyond
the CKM one and no new local operators beyond those present in the
SM, share a useful property \cite{UUT}. Namely, the CKM parameters in
these models extracted from a particular set of data are independent
of the basic functions of section \ref{sec:basic-functions}, they are
universal in this class of models.  Correspondingly there exists a
{\it universal unitarity triangle}.  In particular a determination of
CKM parameters and of this universal unitarity triangle without the
knowledge of supersymmetric parameters is possible.

Indeed using (\ref{eq:xds})  one finds
\begin{equation}\label{107x}
\frac{\vtd}{|V_{ts}|}= 
\xi\sqrt{\frac{m_{B_s}}{m_{B_d}}}
\sqrt{\frac{(\Delta M)_d}{(\Delta M)_s}}\equiv\kappa,
\qquad
\xi = 
\frac{F_{B_s} \sqrt{\hat B_{B_s}}}{F_{B_d} \sqrt{\hat B_{B_d}}}.
\end{equation}
That is the ratio $\kappa$ depends only on the measurable quantities 
$(\Delta M)_{d,s}$, $m_{B_{d,s}}$ and the non-perturbative parameter
$\xi$. Now to an excellent accuracy \cite{BLO}:
\be
\vtd=\vcb\lambda R_t, \qquad R_t=\sqrt{(1-\bar\varrho)^2+\bar\eta^2}
\ee
\be\label{vts1}
\vts=\vcb(1-\frac{1}{2}\lambda^2+\bar\varrho\lambda^2)
\ee
where 
\be
\bar\varrho=\varrho(1-\frac{1}{2}\lambda^2), \qquad
\bar\eta=\eta (1-\frac{1}{2}\lambda^2)
\ee
with $\lambda=0.22$, $\varrho$ and $\eta$ being Wolfenstein
parameters. We note next that through unitarity of the CKM
matrix and the present lower bound on $(\Delta M)_s$ one has
both in the SM and in the MSSM $0\le\bar\varrho\le 0.5$. 
Consequently
$\vts$ deviates from $\vcb$ by at most $2.5\%$. This means
that to a very good accuracy the length of one side of the
unitarity triangle is given by
\be
R_t=\frac{\kappa}{\lambda}
\ee
independently of supersymmetric parameters and $\mt$.
In order to complete the determination of $\bar\varrho$ and $\bar\eta$
one can use $\sin2\beta$ extracted either from the CP asymmetry
in $B_d\to\psi K_S$ or from $K\to\pi\nu\bar\nu$ decays \cite{BBSIN}.
Both extractions are independent to an excellent accuracy of
the supersymmetric parameters and $\mt$. Once $R_t$ and $\sin 2\beta$
have been determined in this manner, $\bar\varrho$ and $\bar\eta$
can be found through \cite{UUT,B95}
\begin{equation}\label{5a}
\bar\eta=\frac{R_t}{\sqrt{2}}\sqrt{\sin 2\beta \cdot r_{-}(\sin 2\beta)}\,,
\quad\quad
\bar\varrho = 1-\bar\eta r_{+}(\sin 2\beta)\,
\end{equation}
where $r_\pm(z)=(1\pm\sqrt{1-z^2})/z$.
In general the calculation of $\bar\varrho$ and $\bar\eta$ from
$R_t$ and $\sin 2\beta$ involves discrete ambiguities.
As described in \cite{UUT,B95}
they can be resolved by using further information, for instance coming from 
$|V_{ub}/V_{cb}|$ and $\varepsilon$, 
so that eventually the solution (\ref{5a})
is singled out.

As an alternative to $\sin 2\beta$ one could use the measurement
of $\sqrt{\bar\varrho^2+\bar\eta^2}$ by means of $\vub$ but this
strategy suffers from hadronic uncertainties in the extraction
of $\vub$.

We observe that provided $(\Delta M)_s$ has been measured and
$\sin 2\beta$ extracted from CP asymmetry in $B_d\to\psi K_S$ or
$K\to\pi\nu\bar\nu$ one can determine the ``true" values of 
$\bar\eta$ and $\bar\varrho$ independently of supersymmetric parameters.
Since $\lambda$ and $\vcb=A\lambda^2$ are determined from tree level
K and B decays they are insensitive to new physics as well. Thus
the full CKM matrix can be determined in this manner. The corresponding
universal unitarity triangle common to the SM and the restricted MSSM
can be found directly from (\ref{5a}). 

Having the universal values of $\bar\varrho$ and $\bar\eta$ at hand one 
can calculate $\varepsilon$,
$\epe$, $(\Delta M)_d$, $(\Delta M)_s$ and branching ratios for
rare decays. As these quantities depend on supersymmetric parameters
the results for the SM and the MSSM will generally differ from
each other and one will be able to find out whether the SM or the MSSM,
if any of these two models, is singled out by the data.
In our opinion the use of the universal unitarity triangle is the
most transparent strategy for distinction between models belonging to
this class. It will certainly play an important role once 
$(\Delta M)_s$ has been measured and $\sin 2\beta$ extracted either from
the CP asymmetry in $B_d^0\to \psi K_S$ or from $K\to\pi\nu\bar\nu$
decays. Other strategies for the determination of the universal unitarity
triangle are discussed in \cite{UUT}.
\subsection{Strategy B: Present}
As $(\Delta M)_s$ and $\sin 2\beta$ from $B_d\to\psi K_S$ or
$K\to\pi\nu\bar\nu$ are presently not available we do not know the
``true" values of the CKM parameters. However, by fitting theoretical
expressions for $\varepsilon$ and $(\Delta M)_d$ to the experimental
data we can find out for which values of CKM parameters the SM and the
MSSM reproduce these data. As this determination involves the basic
functions of section \ref{sec:basic-functions}, the resulting values
of CKM parameters in the MSSM will generally differ from the
corresponding values in the SM.  This difference has to be taken into
account when making predictions for $\epe$ and rare decays branching
ratios.

A different strategy has been adopted in  \cite{JAPAN}. 
These authors have assumed
all CKM parameters in the MSSM to be equal to those in the SM. 
This would be a correct strategy if we knew the universal triangle
but in its absence it could be misleading. Indeed
as a result of this assumption
$(\Delta M)^{\rm MSSM}_d$ in \cite{JAPAN} can
differ from $(\Delta M)^{\rm SM}_d$
by as much as $20\%$ which is well outside the experimental error for
$(\Delta M)_d$ that amounts to $\pm 3\%$. The same comment applies
to the parameter $\varepsilon$ where the experimental error is
even smaller.
This means that either the SM or the MSSM is already ruled out
which of course is not true at present.

In our opinion the only correct procedure at present is to impose
the conditions
\be
(\Delta M)^{\rm SM}_d =(\Delta M)^{\rm MSSM}_d=
(\Delta M)^{\rm exp}_d~,\qquad
(\varepsilon)^{\rm SM} =(\varepsilon)^{\rm MSSM}=
(\varepsilon)^{\rm exp}~.
\ee 
As this conditions are used in our paper,
 our results for the rare
decays branching ratios differ from the ones in \cite{JAPAN}. 

Now, from $(\Delta M)^{\rm SM}_d =(\Delta M)^{\rm MSSM}_d$ it
follows that
\be\label{vtdsusy}
\vtd^2_{\rm MSSM}=\vtd^2_{\rm SM}
\frac{(F_{tt})_{\rm SM}}{(F_{tt})_{\rm MSSM}}
\frac{\eta_B^{\rm SM}}{\eta_B^{\rm MSSM}}~.
\ee
where the last factor takes into account the small difference between QCD
factors in the MSSM and the SM which appears at the two-loop level.
We will set $\eta_B^{\rm MSSM}=\eta_B^{\rm SM}$ in what follows.

As $(F_{tt})_{\rm MSSM}>(F_{tt})_{\rm SM}$ one has
\be
\vtd^2_{\rm MSSM}<\vtd^2_{\rm SM}\,.
\ee
 Similarly,
\be
(\imlt)_{\rm MSSM}<(\imlt)_{\rm SM}, \qquad
\vert(\relt)_{\rm MSSM}\vert<\vert(\relt)_{\rm SM}\vert
\ee
when the constraints from $(\Delta M)_d$ and $\varepsilon$ are
imposed. These hierarchies agree with the recent analysis
in \cite{ALI00}. 

On the other hand $\vts$ can only be modified by supersymmetric
contributions through the change of $\bar\varrho$ in (\ref{vts1}).
As this is an $\ord(\lambda^2)$ effect and $0\le\bar\varrho\le 0.5$
from unitarity and $\vub$ we have within $2.5\%$
\be
\vts_{\rm MSSM}=\vts_{\rm SM}=\vcb.
\ee

These results have the following consequences:

\begin{itemize}
\item
Quantities sensitive to $\vtd$, $\imlt$ and $\relt$ experience
a suppression in the MSSM relative to the SM through the CKM
parameters which may or may not be compensated by the loop
effects relevant for a given process. In particular we have:
\be\label{R1}
\frac{Br(\klpn)_{\rm MSSM}}{Br(\klpn)_{\rm SM}}
=\frac{(\imlt)^2_{\rm MSSM}}{(\imlt)^2_{\rm SM}}
\left[\frac{X^{\nu\bar\nu}_{\rm MSSM}}{X^{\nu\bar\nu}_{\rm SM}}\right]^2~,
\ee
\be\label{R0}
\frac{Br(K_L\to\pi^0 e^-e^+)^{dir}_{\rm MSSM}}
{Br(K_L\to\pi^0 e^-e^+)^{dir}_{\rm SM}}
=\frac{(\imlt)^2_{\rm MSSM}}{(\imlt)^2_{\rm SM}}
\frac{(\tilde y_{7A}^2 + \tilde y_{7V}^2)_{\rm MSSM}}
{(\tilde y_{7A}^2 + \tilde y_{7V}^2)_{\rm SM}}~,
\ee

\be\label{R2}
\frac{(\epe)_{\rm MSSM}}{(\epe)_{\rm SM}}
=\frac{(\imlt)_{\rm MSSM}}{(\imlt)_{\rm SM}}
\frac{F_{\varepsilon'}^{\rm MSSM}}{F_{\varepsilon'}^{\rm SM}}~.
\ee
Moreover in the approximation of neglecting the charm contribution
to $\kpnn$ one finds using (\ref{vtdsusy})
\be\label{R7}
\frac{Br(\kpnn)_{\rm MSSM}}{Br(\kpnn)_{\rm SM}}
=\frac{(F_{tt})_{\rm SM}}{(F_{tt})_{\rm MSSM}}
\left[\frac{X^{\nu\bar\nu}_{\rm MSSM}}{X^{\nu\bar\nu}_{\rm SM}}\right]^2.
\ee
Of course, in our numerical analysis we keep the charm contribution.
\item
Quantities sensitive to $\vts$ depend to a very good approximation 
on supersymmetric effects
only through loop effects relevant for a given process. In particular
we have
\be\label{R4}
\frac{Br(B\to X_s \nu\bar\nu)_{\rm MSSM}}
{Br(B\to X_s \nu\bar\nu)_{\rm SM}}
=\left[\frac{X^{\nu\bar\nu}_{\rm MSSM}}{X^{\nu\bar\nu}_{\rm SM}}\right]^2~,
\ee

\be\label{R5}
\frac{Br(B_s\to \mu\bar\mu)_{\rm MSSM}}
{Br(B_s\to  \mu\bar\mu)_{\rm SM}}
=\left[\frac{Y^{ e\bar e}_{\rm MSSM}}{Y^{e\bar e}_{\rm SM}}\right]^2.
\ee
\item
From (\ref{R1}) and (\ref{R2}) we have
\be\label{R3}
\frac{(\epe)_{\rm MSSM}}{(\epe)_{\rm SM}}
=\sqrt{\frac{Br(\klpn)_{\rm MSSM}}{Br(\klpn)_{\rm SM}}}
\frac{F_{\varepsilon'}^{\rm MSSM}}{F_{\varepsilon'}^{\rm SM}}
\frac{X^{\nu\bar\nu}_{\rm SM}}{X^{\nu\bar\nu}_{\rm MSSM}}~.
\ee
\end{itemize}
These observations should be helpful in understanding the numerical
results presented in section \ref{sec:numerical-analysis}. It should
also be remarked that the formulae (\ref{R1})--(\ref{R3}) are also
valid for strategy A after the $\imlt$ and $F_{tt}$ have been removed
from these equations.

\section{Constraints}
\label{sec:constraints}
\subsection{The $B\to X_s \gamma$ constraint}
The latest data \cite{cleo}
for the inclusive radiative decay rate of 
the $B$ meson are known to constrain severely the accessible supersymmetric
parameter space (see for instance \cite{Erler:1998,bsgconstr}).
 This decay mode is controlled by the 
magnetic penguin operator $Q_7$, whose Wilson coefficient at the weak scale 
can be written as 
\begin{equation}
C_{7\gamma}(M_W)= A^{(\gamma)}_{SM}+A^{(\gamma)}_H+A^{(\gamma)}_\chi,
\end{equation}
where $A^{(\gamma)}_{SM}$, $A^{(\gamma)}_H$, and
$A^{(\gamma)}_\chi$ are the contributions
from $W$, charged Higgs, and chargino loops, respectively.

For a light charged Higgs mass, $A^{(\gamma)}_H$ is large and always 
adds to the SM contribution, driving the branching ratio towards very 
high values. Therefore, only large chargino-stop  contributions 
that interfere destructively with the charged Higgs ones
can accommodate the present data \cite{BARBIERI}. On the other hand,
unless the charged Higgs is very heavy, 
this requires the existence of at least one light chargino and one light stop
and restricts considerably the allowed SUSY parameter space.
In our analysis we therefore include this constraint 
using  the latest theoretical and experimental information.

A complete NLO analysis of $B\to X_s \gamma$
is by now available in the SM \cite{bsgammaSM,GAMB}
and in the 2HDM \cite{GAMB,bsgamma2HDM}. The references to earlier
LO calculations can be found in \cite{AJBLAKE}.
As a result of these  efforts, the error on the 
theoretical prediction coming from uncalculated higher orders is estimated 
to be of the order of 5\%, while large parametric uncertainties
are induced  by the semileptonic BR, used as a normalization,
the bottom and charm masses, the CKM factor, and to a lesser extent 
by the top mass and $\alpha_s$. In the case of the SM and of the type II
2HDM the overall theoretical error is generally around 10\%.
In the numerical analysis we fully implement all the NLO
contributions to SM and 2HDM components including the latest refinements
\cite{CM}.

For what concerns the purely supersymmetric contributions,
there exist no complete NLO calculation. 
However, the case considered in the present
paper (minimal flavour violation) has been studied in
\cite{bsgammaSUSY1,bsgammaSUSY2}, 
where the Wilson coefficients have been expanded
in inverse powers of the gluino and  heavy squark masses. This 
approximation is designed to capture two sources of large NLO effects:
i) if large cancellations between charged Higgs and chargino
contribution occur at LO, they  can be spoiled at NLO ; ii)
the existence of large splitting among supersymmetric particles can lead
to large {\it non-decoupling} logarithms.

In our numerical study we 
employ the results of \cite{bsgammaSUSY1} whenever they are 
applicable: we define a light mass scale $M_l$ as the heavier between
$\tilde m_{t_2}$, $\tilde m_{\chi_1}$,  and a heavy mass scale $M_h$ 
as the minimum of $\tilde m_g$, $\tilde m_{t_1}$ and $\tilde m$; 
if the condition $M_h>2M_l$ is satisfied, the chargino contributions
to the Wilson coefficients are evaluated at NLO, otherwise we use the LO
expressions. Similarly, we employ the NLO expressions only if 
$|\tilde \theta_t|<\pi/10$.

The estimate  of the theoretical error is expected to compensate for this 
different treatment. We adopt a rather conservative approach:
we scan over all the unphysical scales, using
$2.4<\mu_b,\bar{\mu}_b<9.6$ GeV and $40<\mu_W<160$ GeV (notation of 
\cite{GAMB}) and combine in quadrature the parametric uncertainties.
The two extreme values of the BR obtained combining linearly these two 
uncertainties are then compared with the CLEO 95\% C.L. 
$ BR_\gamma >2.0\times 10^{-4} $ and $ BR_\gamma < 4.5 \times 10^{-4}$.
\subsection{Constraints from electroweak precision tests 
}

The good agreement between the SM predictions and the electroweak
 precision  data 
favours those types of new physics for which contributions decouple from
the precision observables. As we also consider light superpartners,
we need to take into account the tight constraints on the
supersymmetric spectrum emerging from this agreement.

In the case of the MSSM, several recent and 
thorough analyses are available in the
literature \cite{Erler:1998,cho}. The most relevant effect 
is due to  the mass splitting of the superpartners, and in
particular of the third generation squarks. Indeed, large splitting between
$\tilde{m}_{b_L}$ and $\tilde{m}_{t_L}$ 
would induce large contributions to the electroweak $\rho$ parameter 
of the same sign of
the standard quadratic top quark term. This universal contribution
enters the $Z^0$ boson couplings and the relation between $M_W$,
$G_\mu$ and $\alpha$ and is therefore significantly constrained by
present data. Following the assumptions of section \ref{sec:mssm}, 
the contribution of the squarks to $\Delta\rho=1-1/\rho$ is given 
at one-loop order by the simple expression \cite{maiani,drees}
\bea
\Delta ^{\tilde q} \rho^{(0)}  &=& \frac{3 G_F}{8 \sqrt{2} \pi^2} \left[ -
\sin^2 \theta_{\tilde{t}} \cos^2 \theta_{\tilde{t}} \,
F\left( m_{\tilde{t}_1}^2,  m_{\tilde{t}_2}^2 \right)
\right. \nonumber \\ 
&& \left. + \cos^2 \theta_{\tilde{t}}\, F\left( m_{\tilde{t}_1}^2,  
m_{\tilde{b}_L}^2 \right) + \sin^2 \theta_{\tilde{t}}\, F\left( 
m_{\tilde{t}_2}^2,  m_{\tilde{b}_L}^2 \right) \right],
\label{drho}
\eea
with the function $F$ given by
\bea
F(x,y)= x+y - \frac{2xy} {x-y} \log \frac{x}{y} \ . 
\nonumber
\eea
Notice that  $\Delta ^{\tilde q} \rho^{(0)}$ in
(\ref{drho}) vanishes for degenerate squarks and   
when one of the squarks becomes very heavy it is proportional to the
square of its mass. Of course, squark contributions are not the only 
supersymmetric
corrections to $\Delta\rho$. However, they are the only
potentially large ones, when the present direct exclusion bounds 
are taken into account, and the other contributions have in general
the same sign \cite{drees}. 

In the cases of the $W$--boson mass and of the effective weak
mixing angle $\sin^2 \Theta_{\rm W}^{\rm eff}$, for example, a
doublet of heavy 
squarks would induce shifts proportional to its contribution to $\rho$,
\be
\delta M_W/M_W \approx 0.72  \,\Delta\rho;
\ \ \ \ 
\delta\sin^2\Theta_{\rm W}^{ eff}\approx - 0.33\, \Delta\rho.
\label{shifts}
\ee
As the relative experimental accuracy for these quantities is 
5$\times10^{-4}$  and $7 \times10^{-4}$, respectively, it is easy to
realize that the present sensitivity to $\Delta\rho$ is at the level of
1$\times 10^{-3}$. 
Of course, a global fit provides a  better 
way to explore this sensitivity.
 The model-independent analysis of \cite{Altarelli}, based on 1998 data, 
finds a value $(3.7\pm 1.1)\, 10^{-3}$ for the parameter $\epsilon_1
\approx \Delta\rho$, where the central value is close to the SM
result, but favours smaller values than in the SM. As $
\Delta\rho_{susy}$ is positive, it is tightly constrained.
The analysis of \cite{erler}, on the other hand, 
includes also the information relative to the restricted Higgs sector of the
MSSM and leads to $\Delta\rho_{susy}= -0.0004^{+0.0017}_{-0.0013}$
at the $2\sigma$ level, which again shows no deviations from the SM.
Another alternative formulation in terms of the parameter $T\approx \alpha
\Delta\rho$  leads to very similar results ($0<T_{susy}<0.2$ at 95\%
CL \cite{Erler:1998,cho}).
We will therefore require  $$\Delta^{\tilde{q}}\rho^{(0)}
< 1.5  \times 10^{-3},$$ which appears
to be  a conservative bound, also taking into account that QCD 
effects tend 
to enhance significantly the squark contributions to $\Delta\rho$ 
\cite{drho2}, and that present data improve on the 1998 results
employed in \cite{Altarelli,erler}, especially on $M_W$. 

\subsection{Direct exclusion bound on $M_h$}
The tree level MSSM relation between the mass of the lightest CP--even
Higgs boson and $M_Z$, $M_h\le M_Z |\cos2 \beta|\le M_Z$, is subject to
very large radiative corrections (a summary of the most recent results
with exhaustive references can be found in \cite{carena}), 
which relax it into a much looser bound
$M_h<135$ GeV. The precise value of $M_h$ depends sensitively on
the supersymmetric parameters and in particular on $\tan\beta$,
the mass of the pseudoscalar, $M_A$ --- which at lowest order 
is linked to the mass of
the charged Higgs by $M_{H^+}^2= M_A^2 + M_W^2$--- and 
the third generation squarks, which give the most important loop
contributions.

At LEP-200 the lightest MSSM neutral Higgs can be produced mainly via
the Higgs-strahlung process $e^+ e^-\to h \, Z$ and the pair
production process $e^+ e^-\to h \, A$. The experimental searches have
so far set a lower bound on the mass of the light neutral Higgs of
about 88 GeV \cite{lephiggs}. As this already excludes large regions
of the supersymmetric parameter space, especially if $\tan\beta$ is
low, it is important to include also this constraint in our analysis.
To this end we employ approximate formulas based on \cite{approx};
unfortunately, they cannot be used for $M_{tLR}> 2 M_{\tilde q}$ and
do not cover part of our parameter space. Therefore, for $M_{tLR}> 2
M_{\tilde q}$ we use the full one-loop expression to
compute the Higgs mass \cite{higgscompl}. We notice that for large
splitting between the stop masses new large logarithms enter the
analysis which may induce relatively large QCD corrections and may
need to be resummed. To the best of our knowledge, this case has not
been considered in the literature. However we believe this does not
affect our main conclusions.

\begin{figure}[tb]
  \begin{center}
     \epsfxsize=10truecm
    \begin{turn}{-90}
    \epsffile{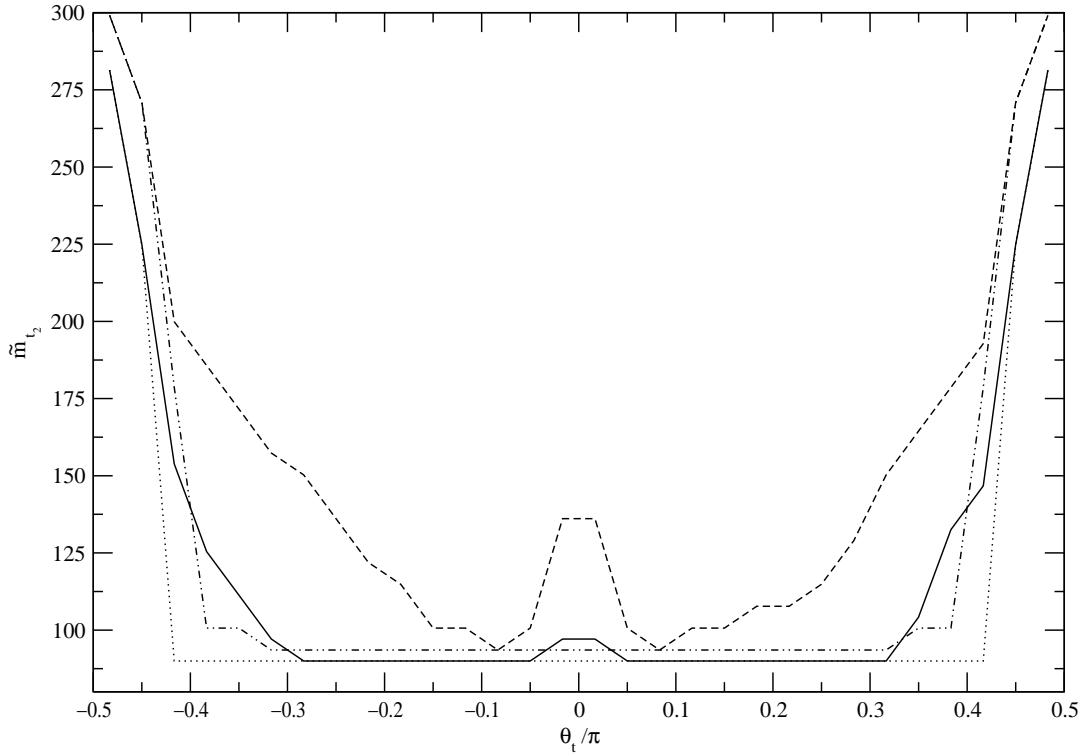}
    \end{turn}   
    \caption{Contour plot in the $\tilde m_{t_2}-\theta_{\tilde t}$
      plane. The regions above the dash-dotted and dotted lines
      contain $95 \%$ and $100 \%$ of the generated points
      respectively, without taking the constraints discussed in
      section \ref{sec:constraints} into account.  The regions above
      the dashed and continuous lines contain $95 \%$ and $100 \%$ of
      the allowed points respectively, once the constraints are taken
      into account.}
    \label{fig:scatter_mst2_thetast}
  \end{center}
\end{figure}

The relevance of the constraints discussed in this section on the SUSY
parameter space can be seen, for example, in
fig.~\ref{fig:scatter_mst2_thetast}, where we present the contour plot
of the light stop mass versus the stop mixing angle in the ranges
described in section \ref{sec:susy-parameter-space}, with and without
the constraints discussed above. Clearly, a large portion of the
region where one of the stops is very light is excluded by the present
constraints.

\section{Numerical Analysis}
\label{sec:numerical-analysis}
\setcounter{equation}{0}
\subsection{Standard Model Parameters}
\label{sec:standard-model-parameters}
In order to make predictions for $\epe$ and rare decays we need the
values of $\IM \lambda_t$ and $\relt$. They can be obtained from the
standard analysis of the unitarity triangle which uses the data for
$\vcb$, $|V_{ub}|$, $\eps$, $(\Delta M)_d$ and $(\Delta M)_s$, where
the last two measure the size of $B^0_{d,s}-\bar B^0_{d,s}$ mixings.
The relevant formulae for $\eps$ and $(\Delta M)_{d,s}$ are given in
sections \ref{subsec:epsformula} and \ref{subsec:BBformula}
respectively.  We also recall that the present lower bound on $(\Delta
M)_s$ together with $(\Delta M)_d$ puts the following constraint on
the ratio $\vtd/|V_{ts}|$:
\begin{equation}\label{107b}
\frac{\vtd}{|V_{ts}|}< 
\xi\sqrt{\frac{m_{B_s}}{m_{B_d}}}
\sqrt{\frac{(\Delta M)_d}{(\Delta M)^{\rm min}_s}}
\end{equation}
with the non-perturbative parameter $\xi$ defined in (\ref{107x}).  As
discussed in section \ref{sec:general-considerations} the constraint in
(\ref{107b}) is valid both in the SM and in the MSSM.

The input parameters needed to perform the analysis of the unitarity 
triangle within the SM are:
\be\label{smpar}
\vcb,\quad |V_{ub}|,\quad \mc,\quad \mt,\quad \hat B_K, 
\quad \sqrt{\hat B_d}F_{B_d}, \quad \xi\,.
\ee
Their values are given in table \ref{tab:inputparams}, where
 $\mt$  refers to the running current top quark mass defined at 
$\mu=\mt^{Pole}$. It corresponds to 
$\mt^{Pole}=174.3\pm 5.1\gev$ measured by CDF and D0.

\begin{table}[tb]
\begin{center}
\begin{tabular}{|c|c|c|}
\hline
{\bf Quantity} & {\bf Central} & {\bf Error}  \\
\hline
$|V_{cb}|$ & 0.040 & $\pm 0.002$   \\
$|V_{ub}|$ & $3.56\cdot 10^{-3}$ & $\pm 0.56\cdot 10^{-3} $ \\
$\mc$ & $1.3\gev$ & 0.1 \gev\\
$\mt$ & $165\gev$ & $\pm 5\gev$  \\
$\hat B_K$ & 0.80 & $\pm 0.15$   \\
$\sqrt{B_d} F_{B_{d}}$ & $200\mev$ & $\pm 40\mev$ \\
$(\Delta M)_d$ & $0.471~\mbox{ps}^{-1}$ & $\pm 0.016~\mbox{ps}^{-1}$ 
\\
$(\Delta M)_s$ & $>14.6~\mbox{ps}^{-1}$ & $ 95\% {\rm C.L.}$ 
\\
$\xi$ & $1.16$ & $\pm 0.07$  \\
$\Lms^{(4)}$ & $340 \mev$ & $\pm 50\mev$  \\
$\ms(\mc)$ & $130\mev$    & $\pm 25\mev$ \\
$\bsi $ & 1.0 & $\pm 0.3$ \\
$\bei $ & 0.8 & $\pm 0.2$ \\
\hline
\end{tabular}
\end{center}
\caption[]{Collection of SM input parameters.
\label{tab:inputparams}}
\end{table}

It should be emphasized that $\hat B_K$, $\sqrt{\hat B_d}F_{B_d}$ 
and $\xi$
being obtained from the calculations of hadronic matrix elements in
QCD are unaffected by the presence of supersymmetry. The extraction
of the remaining parameters in (\ref{smpar}) from the data could
in principle be affected by supersymmetric contributions but
at least within MSSM such contributions are safely negligible.

The additional input parameters required to perform the analysis
of $\epe$ within the SM are
\be\label{epepar}
\Lms^{(4)},\quad \ms(\mc),\quad \bsi,\quad \bei
\ee
Their values are given in table \ref{tab:inputparams}, where
the values for $\Lms^{(4)}$ and $\ms(\mc)$ correspond roughly to 
$\alpha_s(\mz)=0.119\pm 0.003$, and $\ms(2~\gev)=(110\pm20)\mev$
respectively.

Again the non-perturbative parameters $\bsi$ and $\bei$
are unaffected by the presence of supersymmetry. The extraction
of $\Lms^{(4)}$ and $\ms(\mc)$ from the data could
in principle be affected by supersymmetric contributions but
at least within the MSSM such contributions are safely negligible.

In summary the values of the input parameters given in table
\ref{tab:inputparams} are common to the SM and MSSM. On the other hand
the resulting values of $\imlt$, $\relt$, $\epe$ and branching ratios
for rare decays will be generally different in these two models due to
new supersymmetric contributions in the one-loop diagrams as
summarized in sections \ref{sec:basic-functions} and
\ref{sec:compendium}.  We follow here the strategy B of section
\ref{sec:general-considerations} as the universal unitarity triangle
is unknown at present.

The values of the parameters in table \ref{tab:inputparams} are
precisely the ones used in \cite{EP99} except for the lower
bound on $(\Delta M)_s$ which has been improved since then \cite{LEPMS}.
The relevant references to table \ref{tab:inputparams} can be found
in \cite{EP99}.

The only controversial values in table \ref{tab:inputparams} are
the ones for $\bsi$, $\bei$ and $\ms$. Other authors would possibly
use different values for these parameters. As one of the main purposes
of the present paper is to investigate the impact of supersymmetric
contributions on our $\epe$ analysis within the SM
\cite{EP99}, we will keep the values of all the SM
input parameters as in \cite{EP99}. Reviews of the values of
$\bsi$, $\bei$ and $\ms$ used by other authors can be found in
\cite{EP99,BUJA}. We will briefly comment at the end of this
section how our analysis would change had we used different
values for $\bsi$ and $\bei$.

\subsection{Supersymmetric Parameters}
\label{sec:susy-parameter-space}
As discussed in section \ref{subsec:the-model}, we will choose the
following supersymmetric  
input parameters:
\be\label{susypar}
\tilde m_{t_1}, \quad \tilde m_{t_2},\quad \theta_{\tilde t},\quad
\tilde m_{\chi_1}, \quad \tan\beta,\quad m_{H^\pm}, \quad \mu, \quad
m_{\tilde \ell},
\ee
where $\tilde m_{t_1}$ and $\tilde m_{t_2}$ are the heavy and light
stop mass respectively, $\theta_{\tilde t}$ is the mixing
angle in the stop mass matrix,
$\tilde m_{\chi_1}$ the light chargino mass, $\tan\beta$ the ratio
of the two Higgs vacuum expectation values, $m_{H^\pm}$ the
charged Higgs mass, $\mu$ the Higgs superfield mixing parameter and
$m_{\tilde \ell}$ is the common slepton mass entering the expressions
for rare K and B decays. We now discuss the ranges in which we
randomly generate with flat distributions the parameters in
eq.~(\ref{susypar}).

The region of SUSY parameter space in which we expect SUSY
contributions to $\varepsilon^\prime/\varepsilon$ to be maximal (in
absolute value) corresponds to the case of a light stop and chargino.
We therefore allow for the lighter stop and chargino to be as light as
allowed by present constraints,
\begin{equation}
  \label{eq:3}
   90 \,\mathrm{GeV} < \tilde m_{\chi_1} < 250 
   \,\mathrm{GeV}, \quad 90 \,\mathrm{GeV} < \tilde m_{t_2} < 800 
   \,\mathrm{GeV}.
\end{equation}
Concerning the stop mixing angle, we generate it in the range $-\pi/2 <
\theta_{\tilde t} < \pi/2$. The case of a very light and
right-handed $\tilde t_2$ requires $\tilde m^2_{\tilde t_R} < 0$. This
might however induce charge and colour breaking minima in the
potential. To avoid this, following ref.~\cite{carena2} we impose the
bound
\begin{equation}
  \label{eq:mstopr}
  \tilde m^2_{\tilde t_R} > - \left(\frac{M_h^2 (v_1^2+v_2^2)
  g_s}{12}\right)^{\frac{1}{2}}\,, 
\end{equation}
which for a typical Higgs mass $M_h\sim 90$ GeV amounts to $\tilde
m^2_{\tilde t_R} \gsim - (90$ GeV$)^2$. We also impose $\tilde m >
250$: Since minimal flavour violation requires the soft breaking
left-handed squark mass to be universal, this implies $\tilde m_{t_2}
\gsim 300$ GeV for $\theta_{\tilde t} \sim \pm \pi/2$. 

We vary the charged Higgs mass in the range $100 \,\mathrm{GeV} <
m_{H^\pm} < 1 \, \mathrm{TeV}$, and choose $\vert \mu \vert < 500
\,\mathrm{GeV}$. In this work, we do not analyze possible effects in
the large $\tan \beta$ regime, where additional operators can arise.
We can therefore limit ourselves to the region $1.2 < \tan \beta < 6$,
since the dependence on $\tan \beta$ becomes very weak already for
$\tan \beta >3$, until one enters the large $\tan \beta$ regime.
Finally, we impose $\tilde m_{t_1} < 1 \, \mathrm{TeV}$ and $90
\,\mathrm{GeV} < \tilde m_\ell < 500 \,\mathrm{GeV}$. We do not
enforce GUT relations between SUSY parameters since we are considering
a more general class of theories described in terms of the SUSY
parameters at the electroweak scale, with no reference to the
structure of the theory at higher scales.  The NLO calculation adopted
for $BR(b \to s \gamma)$ introduces a dependence also on the gluino
mass $\tilde m_g$, which we choose to vary in the range $250
\,\mathrm{GeV} < \tilde m_g < 1 \, \mathrm{TeV}$.

We scan the parameter space specified above and impose the constraints
discussed in section \ref{sec:constraints}. 
Then for the points allowed by these constraints
we perform the determination of $\IM \lambda_t$ and $\RE\lambda_t$
from $\Delta F=2$ processes and then compute
$\varepsilon^\prime/\varepsilon$ and branching ratios for rare decays. 
We provide the details of this computation below.

The ranges for supersymmetric input parameters are summarized in
table \ref{tab:susyinput}.

\begin{table}[tb]
\begin{center}
\begin{tabular}{|c|c|c|}
\hline
{\bf Quantity} & {\bf Min} & {\bf Max}  \\
\hline
$\tilde m_{t_1}$ & $90 \,\mathrm{GeV} $&$ 1 \, \mathrm{TeV}$ \\
$\tilde m_{t_2}$ & $90 \,\mathrm{GeV} $&$ 800 \,\mathrm{GeV}$ \\
$\theta_{\tilde t}$ & $-\pi/2 $&$ \pi/2$\\
$\tilde m_{\chi_1}$ & $90 \,\mathrm{GeV} $&$ 250 \,\mathrm{GeV}$  \\
$\tilde m_{g}$ & $250 \,\mathrm{GeV} $&$ 1 \, \mathrm{TeV}$  \\
$\tan\beta$ & $1.2 $&$ 6$  \\
$m_{H^\pm} $ & $100 \,\mathrm{GeV} $&$ 1 \, \mathrm{TeV}$ \\
$\mu$ & $-500 \,\mathrm{GeV} $&$ 500 \,\mathrm{GeV}$\\
$\tilde m_{\ell}$ & $90 \,\mathrm{GeV} $&$ 500 \,\mathrm{GeV}$\\
\hline
\end{tabular}
\end{center}
\caption[]{Collection of supersymmetric input parameters.
\label{tab:susyinput}}
\end{table}

\subsection{Strategy}
The uncertainties in the input parameters of table \ref{tab:inputparams}
screen considerably the effects of supersymmetric contributions.
In order to study various trends and patterns of supersymmetric effects
in $\imlt$, $\relt$, $\epe$ and rare decays we proceed as follows:

{\bf Step 1:}

We set the SM parameters to 
their central values in table \ref{tab:inputparams} and, using a
random generation of $\sim 6\cdot 10^{6}$ points, calculate the
ranges for the ratios
\be\label{kratios}
K(F_r)=\frac{(F_r)_{\rm MSSM}}{(F_r)_{\rm SM}},
\qquad  F_r=X,Y,Z,E,S,X^{(\nu\bar\nu)},Y^{(e\bar e)}
\ee
with $(F_r)_{\rm SM}$ and $(F_r)_{\rm MSSM}$ representing the SM and
total contributions to the functions in question respectively.
$S\equiv F_{tt}$ with $F_{tt}$ given in (\ref{tt}).
These ratios are of interest for both strategies of section
\ref{sec:general-considerations}.
In order to investigate the impact of various constraints on SUSY
parameters we consider the following cases:
\begin{enumerate}
\item
Only the constraints from $\varepsilon$, $B^0_{d,s}-\bar B^0_{d,s}$-
mixings and direct bounds on masses of squarks, sleptons, charginos and
charged Higgs particles are imposed. 
\item
The additional constraint from $B\to X_s\gamma$ is imposed.
\item
The additional constraint from  $\Delta\varrho$ is imposed 
 without any constraint  from $B\to X_s\gamma$.
\item The additional constraint from the lower bound on the neutral
  Higgs mass is imposed without any constraint from $B\to X_s\gamma$
  and $\Delta\varrho$.
\item
All constraints are taken into account.
\end{enumerate}
This step allows us to identify the dominant supersymmetric contributions.

{\bf Step 2:}

We repeat the analysis of Step 1 calculating this time the ranges for
the ratios
\be\label{Rratios}
T(\epe)=\frac{(\epe)_{\rm MSSM}}{(\epe)_{\rm SM}}
\ee
and analogous ratios for  $\imlt$, $\relt$, and the branching ratios
for rare decays. 

We also investigate for which sets of supersymmetric parameters
the maximal and minimal values of these ratios are obtained.

{\bf Step 3:}

We repeat the analysis of Step 2 in the case of $\epe$, investigating
the dependence of $T(\epe)$ on $B_6^{(1/2)}$.

\subsection{Results of Step 1}
Our analysis shows that, taking only the constraints (1) into account,
supersymmetric contributions to the box functions $B^{(u)}$, $B^{(d)}$
and $B^{(\nu\bar\nu)}$ amount respectively to at most $4\%$, $7\%$ and
$11\%$ of their SM values and are suppressed below $3\%$, $6\%$ and
$11\%$ after all constraints are taken into account. These effects are
only relevant in the case of the function $X$ where the box function
enters with an additional factor 4.  The supersymmetric contributions
to the functions $\vert B^{(e \bar e)} \vert$, $C$, $\vert D \vert$,
$E$ and $S$ can be as high as $+34\%$, $+51\%$, $+93\%$, $+107\%$
and $+159\%$ respectively if no constraints from $B\to X_s\gamma$,
$\Delta\varrho$ and $(M_{h})_{min}$ are imposed. While $S$ is
always enhanced, $C$, $\vert D \vert$ and $E$ can be suppressed up to
$24\%$, $7\%$ and $17\%$ respectively.  The constraints (2--5) turn out to
have only a minor impact on the functions $B^{(e \bar e)}$, $D$, $E$
and $S$.  On the other hand the maximal enhancement of $C$ is reduced
to $29\%$.

The function $E$ is absent in $\kpnn$, $\klpn$, $\bnunu$ and $\bmumu$,
and plays only a minor role in $\epe$ and a negligible role in
$\klee$.  Consequently its modification is immaterial for our
analysis.  On the other hand the enhancement of $S$ can have
considerable impact on the values of $\imlt$ and $\relt$ as we will
see below. Similarly the supersymmetric effects in the functions $C$
and $D$ are sizable, although these effects are less visible in the
functions $X$, $Y$, $Z$ due to various cancellations between box and
penguin contributions.

In table~\ref{tab:K} we show the minimal and maximal values of the
ratios $K(F_r)$ defined in (\ref{kratios}) in the case of no
constraints from $B\to X_s\gamma$, $\Delta\varrho$ and
$(M_{h})_{min}$ and after the imposition of these constraints. The
anatomy of these three constraints is shown in table~\ref{tab:CK}. In
the case of the functions $E$ and $S$ we observe the features just
discussed. The supersymmetric effects in the functions $X$, $Y$ and
$Z$ can be at most $\pm 15\%$, $\pm 23\%$ and $\pm 35\%$ respectively
after all constraints have been imposed.  This is related to the fact
that while the function $C$ is positive, the function $D$ is negative.
Consequently the observed enhancement of $\vert D\vert$ results in a
suppression of $Z$. This suppression is reduced by the factor $1/4$
with which the function $D$ enters the function $Z$ and by the
enhancement of $C$, which for certain values of supersymmetric
parameters can overcompensate the suppression of $Z$ through $D$ so
that an enhancement of $Z$ is possible.

\begin{table}[tb]
\begin{center}
\begin{tabular}{|c|c|c|c|c|}
\hline
& \multicolumn{2}{c|}{{\rm No Constraints}} &
  \multicolumn{2}{c|}{{\rm All Constraints}} \\
\hline
$K(F_r)$ & {\bf min} & {\bf max} & {\bf min} & {\bf max} \\
\hline
$K(B^{(u)})$ & $0.96$ & $1.00$ & $0.97$ & $1.00$ \\
$K(B^{(d)})$ & $0.93$ & $1.01$ & $0.94$ & $1.01$ \\
$K(B^{(\nu\bar\nu)})$ & $0.89$ & $1.01$ & $0.89$ & $1.01$ \\
$K(B^{(e\bar e)})$ & $0.94$ & $1.34$ & $0.96$ & $1.34$ \\
$K(C)$ & $0.76$ & $1.51$ & $0.76$ & $1.29$ \\
$K(D)$ & $0.93$ & $1.93$ & $0.95$ & $1.87$ \\
$K(X)$ & $0.87$ & $1.26$ & $0.87$ & $1.15$ \\
$K(Y)$ & $0.81$ & $1.41$ & $0.81$ & $1.23$ \\
$K(Z)$ & $0.68$ & $1.65$ & $0.70$ & $1.35$ \\
$K(E)$ & $0.83$ & $2.07$ & $0.83$ & $1.85$ \\
$K(S)$ & $1.01$ & $2.59$ & $1.01$ & $2.13$ \\
$K(X^{(\nu\bar\nu)})$ & $0.83$ & $1.25$ & $0.85$ & $1.15$ \\
$K(Y^{(e\bar e)})$    & $0.82$ & $1.41$ & $0.82$ & $1.24$ \\
$K(F_{\varepsilon'})$    & $0.59$ & $1.27$ & $0.76$ & $1.23$ \\
\hline
\end{tabular}
\end{center}
\caption[]{The minimal and maximal values of the the ratios $K(F_r)$
without constraints and with all constraints taken into account.
\label{tab:K}}
\end{table}

\begin{table}[tb]
\begin{center}
\begin{tabular}{|c|c|c|c|c|c|c|}
\hline
& \multicolumn{2}{c|}{$B\to X_s \gamma$ only} &
  \multicolumn{2}{c|}{$\Delta\varrho$ only} &
  \multicolumn{2}{c|}{$(M_{h})_{min}$ only} \\
\hline
$K(F_r)$ & {\bf min} & {\bf max} & {\bf min} & {\bf max}
 & {\bf min} & {\bf max} \\
\hline
$K(B^{(u)})$ & $0.96$ & $1.00$ &$0.96$ & $1.00$ & $0.96$ & $1.00$ \\
$K(B^{(d)})$ & $0.93$ & $1.01$ & $0.94$ & $1.01$ &$0.93$ & $1.01$ \\
$K(B^{(\nu\bar\nu)})$ & $0.89$ & $1.01$ & $0.89$ &$1.01$ & $0.89$ & $1.01$ \\
$K(B^{(e\bar e)})$ & $0.95$ & $1.34$ & $0.94$ &$1.34$ & $0.95$ & $1.34$ \\
$K(C)$ & $0.76$ & $1.51$ & $0.76$ & $1.39$ &$0.76$ & $1.51$ \\
$K(D)$ & $0.93$ & $1.93$ & $0.93$ & $1.93$ &$0.94$ & $1.87$ \\
$K(X)$   &$0.87$ & $1.26$ & $0.87$ & $1.20$ & $0.87$ &
 $1.26$ \\
$K(Y)$   &$0.81$ & $1.41$ & $0.81$ & $1.31$ & $0.81$ &
 $1.41$  \\
$K(Z)$   &$0.68$ & $1.65$ & $0.68$ & $1.47$ & $0.70$ &
 $1.65$  \\
$K(E)$   &$0.83$ & $2.07$ & $0.83$ & $2.07$ & $0.83$ & 
$1.85$  \\
$K(S)$   &$1.01$ & $2.59$ & $1.01$ & $2.59$ & $1.01$ &
 $2.18$  \\
$K(X^{(\nu\bar\nu)})$ & $0.83$ & $1.25$ & $0.83$ & $1.20$ 
& $0.85$ & $1.25$ \\
$K(Y^{(e\bar e)})$    & $0.82$ & $1.41$ & $0.82$ & $1.31$
 & $0.82$ & $1.41$ \\
$K(F_{\varepsilon'})$ & $0.61$ & $1.27$ & $0.68$ & $1.27$
 & $0.59$ & $1.23$ \\
\hline
\end{tabular}
\end{center}
\caption[]{The anatomy of various constraints on the minimal and maximal 
values of the the ratios $K(F_r)$.
\label{tab:CK}}
\end{table}

It should be stressed that in the two Higgs doublet model
the function $Z$ is always enhanced through charged Higgs
effects. The possible suppression of $Z$ in the MSSM comes then
from chargino exchanges that are particularly important
in the photon penguin function $D$.

\begin{figure}[tb]
  \begin{center}
    \epsfxsize=11.5truecm
    \begin{turn}{-90}
    \epsffile{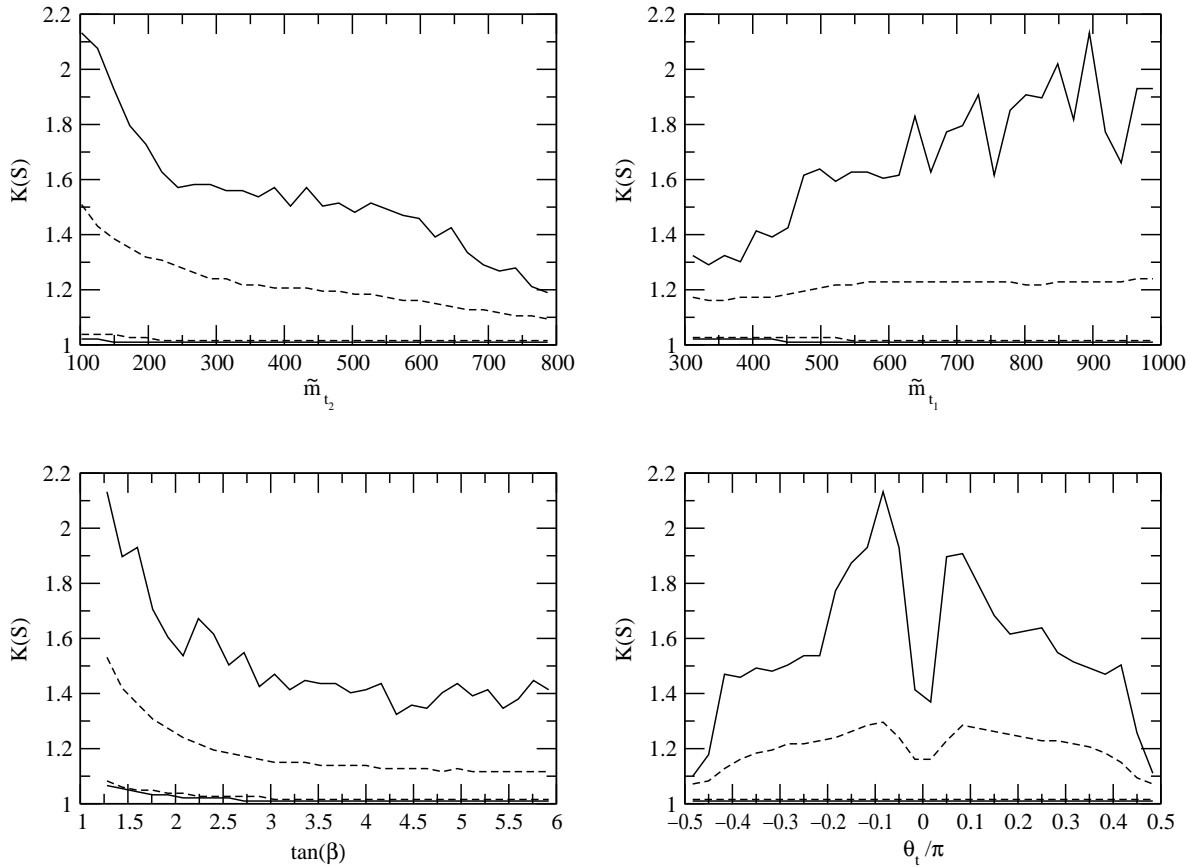}
    \end{turn}
    \caption{Contour plots of $K(S)$ as a function of $\tilde m_{t_2}$,
      $\tilde m_{t_1}$, $\tan \beta$ and $\theta_{\tilde t}$, including
      all the constraints discussed in section
      \ref{sec:constraints}. The dashed and continuous lines contain
      $95\%$ and $100\%$ of 
      the allowed points respectively.} 
    \label{fig:scatters}
  \end{center}
\end{figure}

\begin{figure}[tb]
  \begin{center}
    \epsfxsize=11.5truecm
    \begin{turn}{-90}
    \epsffile{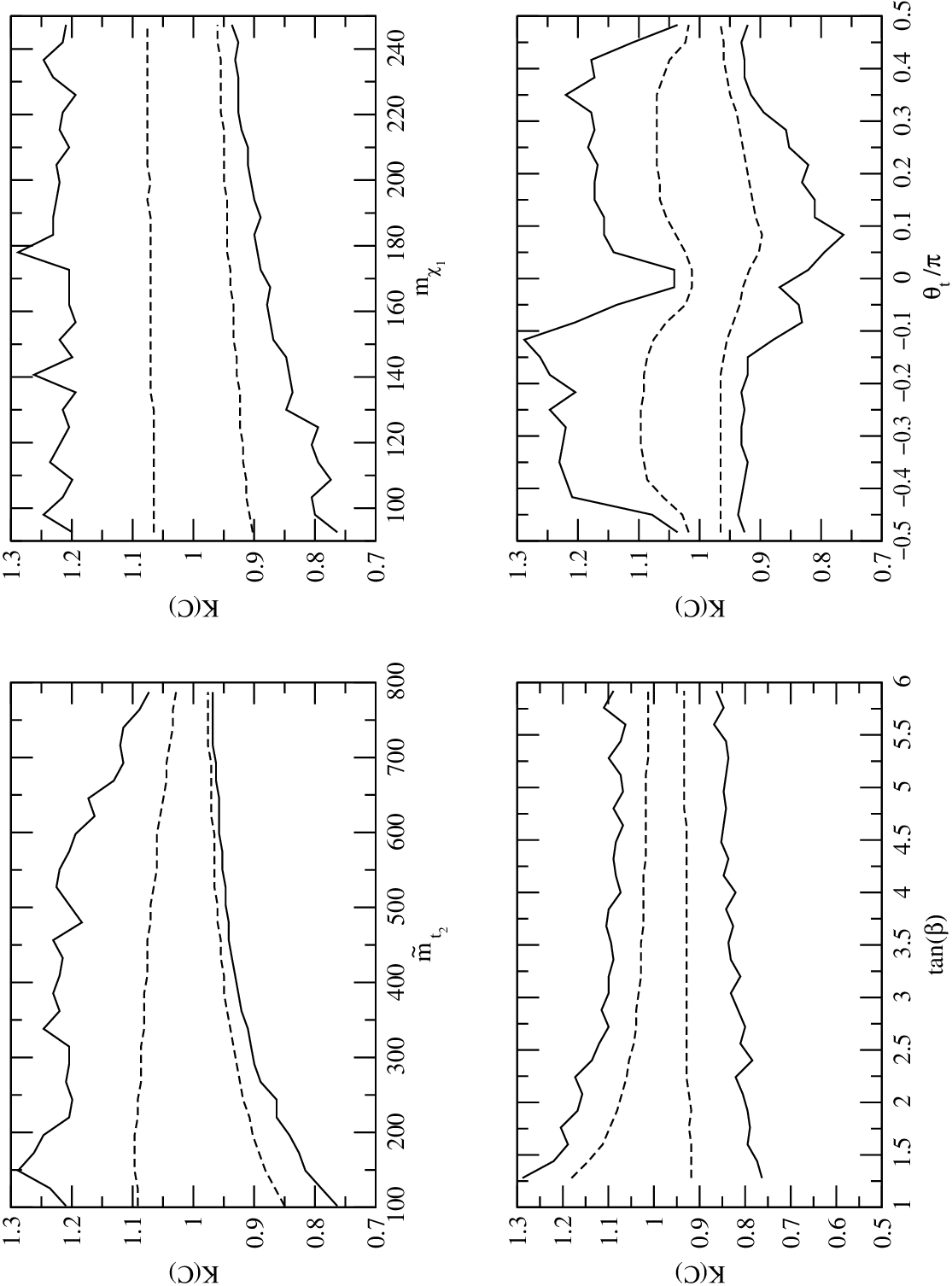}
    \end{turn}    
    \caption{Contour plots of $K(C)$ as a function of $\tilde m_{t_2}$,
      $\tilde m_{t_1}$, $\tan \beta$ and $\theta_{\tilde t}$, including
      all the constraints discussed in section \ref{sec:constraints}.
      The dashed and continuous lines contain $95\%$ and $100\%$ of
      the allowed points respectively.}
    \label{fig:scatterc}
  \end{center}
\end{figure}

\begin{figure}[tb]
  \begin{center}
    \epsfxsize=11.5truecm
    \begin{turn}{-90}
    \epsffile{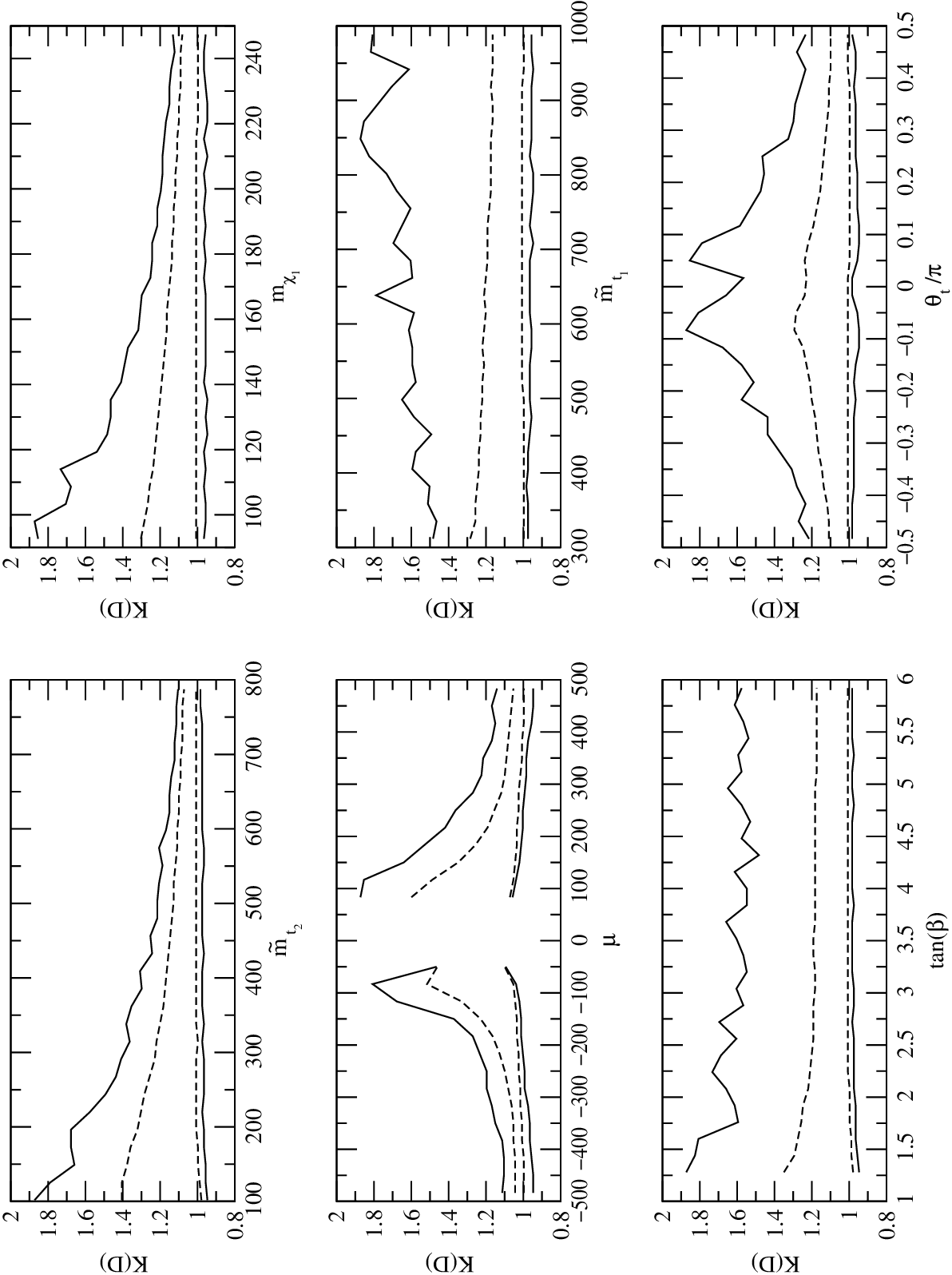}
    \end{turn}    
    \caption{Contour plots of $K(D)$ as a function of $\tilde m_{t_2}$,
      $\tilde m_{\chi_1}$, $\mu$, $\tilde m_{t_1}$, $\tan \beta$ and
      $\theta_{\tilde t}$, including 
      all the constraints discussed in section \ref{sec:constraints}.
      The dashed and continuous lines contain $95\%$ and $100\%$ of
      the allowed points respectively.}
    \label{fig:scatterd}
  \end{center}
\end{figure}

\begin{figure}[tb]
  \begin{center}
    \epsfxsize=11.5truecm
     \begin{turn}{-90}
    \epsffile{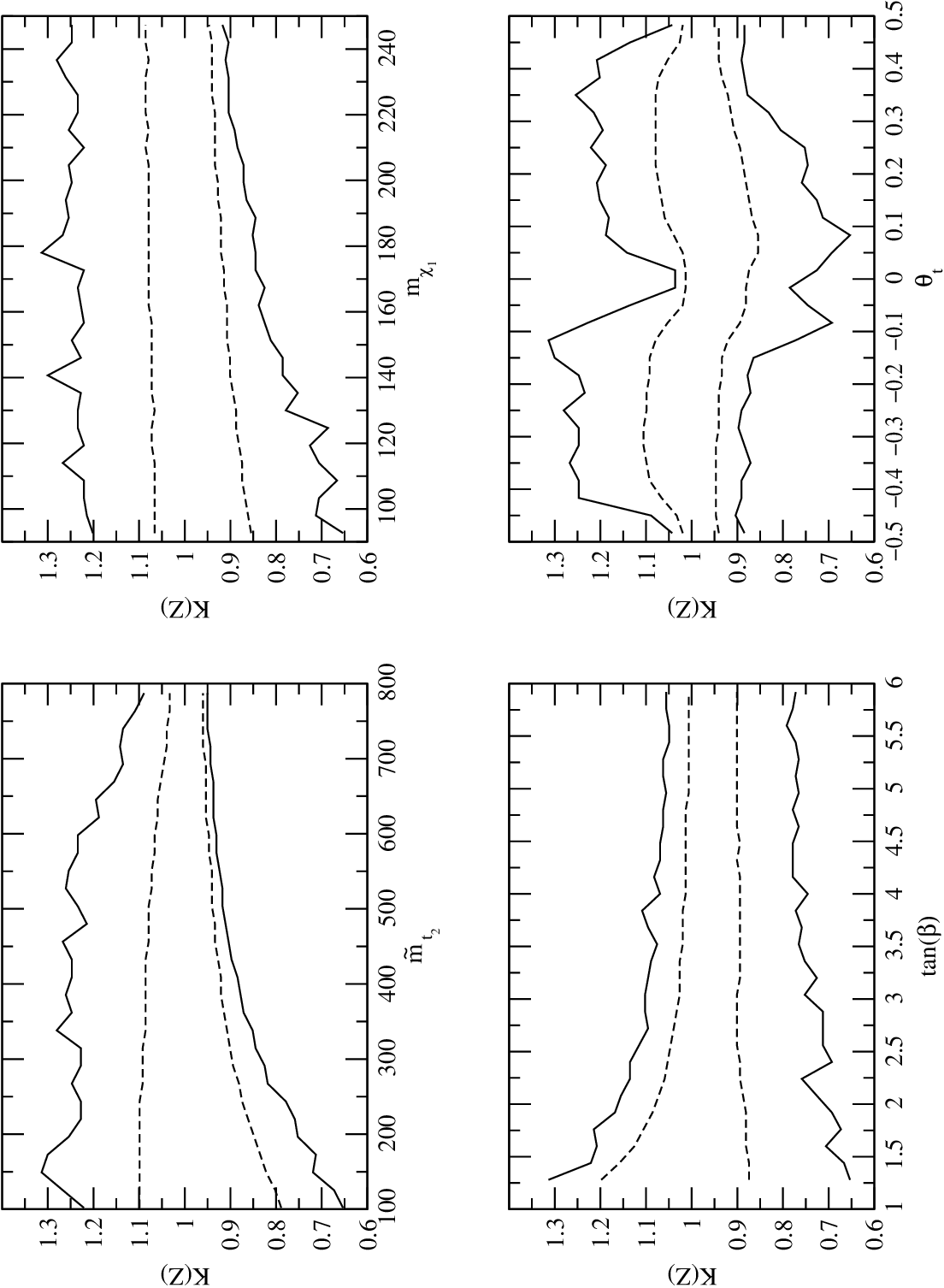}
    \end{turn}       
    \caption{Contour plots of $K(Z)$ as a function of $\tilde m_{t_2}$,
      $\tilde m_{t_1}$, $\tan \beta$ and $\theta_{\tilde t}$, including
      all the constraints discussed in section \ref{sec:constraints}.
      The dashed and continuous lines contain $95\%$ and $100\%$ of
      the allowed points respectively.}
    \label{fig:scatterz}
  \end{center}
\end{figure}

Let us now identify the region of SUSY parameter space where the
maximal deviations from the SM are possible. The function $S$ takes
its largest values for a light, mainly right-handed stop, with a large
splitting between the stops, and small $\tan \beta$ (see
fig.~\ref{fig:scatters}). The function $C$
reaches its maximal values for small $\tan \beta$ and large stop
mixing, while the minimal values are attained for a light, mainly
right-handed stop and light chargino (see fig.~\ref{fig:scatterc}). 
The function $\vert D \vert$ is
almost always larger than in the SM, and reaches its maximum for a
light, mainly right-handed stop, a light chargino and small $\mu$ (see
fig.~\ref{fig:scatterd}).
Therefore, $Z$ is enhanced with respect to the SM for small $\tan
\beta$ and large stop mixing, and suppressed for a
light, mainly right-handed stop and light chargino (See
fig.~\ref{fig:scatterz}).


As seen in table \ref{tab:CK} the constraints from $\Delta\varrho$ and
the lower bound on $H^0$ mass are slightly more effective than the
constraint from $B\to X_s\gamma$. Yet the comparison of tables
\ref{tab:K} and \ref{tab:CK} shows that the inclusion of all bounds
simultaneously eliminates a range of possible values for $X$, $Y$ and
$Z$ which are allowed if each constraint is imposed separately.

\begin{figure}[tb]
  \begin{center}
    \epsfxsize=11.5truecm
     \begin{turn}{-90}
    \epsffile{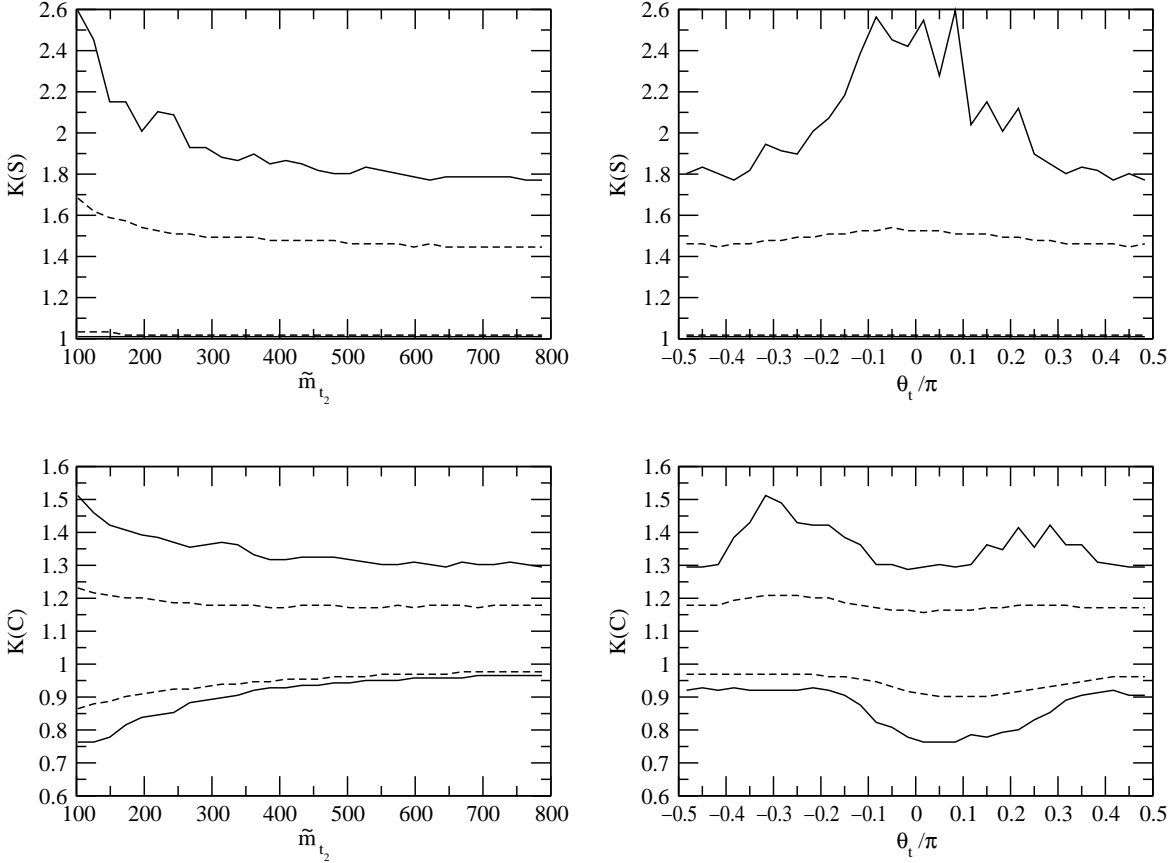}
    \end{turn}           
    \caption{Contour plots of $K(S)$ and $K(C)$ as a function of
      $\tilde m_{t_2}$ and $\theta_{\tilde t}$, without imposing the
      constraints discussed in section \ref{sec:constraints}. The
      dashed and continuous lines contain $95\%$ and $100\%$ of the
      generated points respectively.}
    \label{fig:scconstr}
  \end{center}
\end{figure}

To evaluate the effect of the bounds discussed in section
\ref{sec:constraints}, we show in fig.~\ref{fig:scconstr} the ranges
for $K(S)$ and $K(C)$ without applying the constraints: the comparison
with figs.~\ref{fig:scatters} and \ref{fig:scatterc} clearly shows the
impact of the bounds. 

From the above considerations it is clear that any future increase in
the lower bounds on $M_h$ and $\tan \beta$ will considerably reduce
the possible enhancements of $S$ and $Z$, and therefore, as we shall
see in the following, the possible suppression of $\epe$.

\subsection{Results of Step 2}
In table~\ref{tab:R} we show the minimal and maximal values of the
ratios $T$ defined in (\ref{Rratios}) in the case of no constraints
from $B\to X_s\gamma$, $\Delta\varrho$ and $(M_{h})_{min}$ and after
the imposition of these bounds. The anatomy of these three constraints
is shown in table~\ref{tab:CR}.  We make the following observations,
taking all the constraints into account:

\begin{itemize}
\item The extracted $\imlt$ and $\vert\relt\vert$ are suppressed by
  supersymmetric contributions. This is related to the enhancement of
  $F_{tt}$ in the MSSM.
\item $\epe$ can be enhanced by at most $7\%$ but can be suppressed
  even by a factor of $2$. This pattern agrees with the one found in
  \cite{GG} although the supersymmetric effects in $\epe$ found here
  are smaller than in \cite{GG} due to the tighter constraints on
  supersymmetric parameters imposed in the present analysis. We will
  elaborate on these results below.
\item The branching ratios for $\kpnn$, $\klpn$ and $\klee$ are
  basically suppressed by supersymmetric contributions relative to the
  SM expectations.  The suppression of $Br(\kpnn)$ can be at most
  $35\%$. The corresponding suppressions in $\klpn$ and $\klee$ can be
  as high as a factor of $2.5$ and $2$ respectively.  As seen in
  (\ref{R1}), (\ref{R0}) and (\ref{R7}) and tables~\ref{tab:K} and
  \ref{tab:R}, these suppressions are related to the suppressions of
  $\imlt$ and $\vert \relt\vert $ that cannot be compensated by
  possible enhancements of the functions $X$, $Y$ and $Z$.
\item As expected on the basis of (\ref{R4}) and (\ref{R5}) the
  pattern of supersymmetric effects is different for the rare decays
  $\bnunu$ and $\bmumu$.  Here possible enhancements and suppressions
  can be directly calculated using (\ref{R4}), (\ref{R5}) and table
  \ref{tab:K}. The supersymmetric effects in $Br(\bnunu)$ can be at
  most $\pm 34\%$. $Br(\bmumu)$ can be enhanced up to $53\%$ and
  suppressed up to $32\%$.
\end{itemize}

\begin{table}[tb]
\begin{center}
\begin{tabular}{|c|c|c|c|c|}
\hline
& \multicolumn{2}{c|}{{\rm No Constraints}} &
  \multicolumn{2}{c|}{{\rm All Constraints}} \\
\hline
$T $ & {\bf min} & {\bf max} & {\bf min} & {\bf max} \\
\hline
$T(\imlt)$ & $0.57$ & $1.00$ & $0.66$ & $1.00$    \\
$T(\relt)$ & $0.78$ & $1.00$ & $0.81$ & $1.00$     \\
$T(\epe)$ & $0.42$ & $1.07$ & $0.53$ & $1.07$ \\
$T(\kpnn)$ & $0.59$ & $1.09$ & $0.65$ & $1.02$  \\
$T(\klpn)$ & $0.28$ & $1.12$ & $0.41$ & $1.03$     \\
$T(\klee)$ & $0.33$ & $1.10$ & $0.48$ & $1.10$  \\
$T(\bnunu)$ & $0.70$ & $1.60$ & $0.73$ & $1.34$     \\
$T(\bmumu)$ & $0.68$ & $1.99$ & $0.68$ & $1.53$     \\
\hline
\end{tabular}
\end{center}
\caption[]{The minimal and maximal values of the the ratios $T$
without constraints and with all constraints taken into account.
\label{tab:R}}
\end{table}

\begin{table}[tb]
\begin{center}
\begin{tabular}{|c|c|c|c|c|c|c|}
\hline
& \multicolumn{2}{c|}{$B\to X_s \gamma$ only} &
  \multicolumn{2}{c|}{$\Delta\varrho$ only} &
  \multicolumn{2}{c|}{$(M_{h})_{min}$ only} \\
\hline
$T$ & {\bf min} & {\bf max} & {\bf min} & {\bf max}
 & {\bf min} & {\bf max} \\
\hline
$T(\imlt)$      & $0.57$ & $1.00$ & $0.57$ & $1.00$ & $0.64$ & $1.00$ \\
$T(\relt)$      & $0.78$ & $1.00$ & $0.78$ & $1.00$ & $0.80$ & $1.00$ \\
$T(\epe)$       & $0.42$ & $1.07$ & $0.46$ & $1.07$ & $0.42$ & $1.07$  \\
$T(\kpnn)$    & $0.59$ & $1.08$ & $0.59$ & $1.02$ & $0.65$ & $1.09$\\
$T(\klpn)$    & $0.28$ & $1.12$ & $0.28$ & $1.03$ & $0.41$ & $1.12$\\
$T(\klee)$      & $0.33$ & $1.10$ & $0.33$ & $1.10$ & $0.48$ & $1.10$\\
$T(\bnunu)$      & $0.70$ & $1.60$ & $0.70$ & $1.46$ & $0.73$ & $1.60$\\
$T(\bmumu)$ & $0.68$ & $1.99$ & $0.68$ & $1.73$ & $0.68$ & $1.99$ \\
\hline
\end{tabular}
\end{center}
\caption[]{The anatomy of various constraints on the minimal and maximal 
values of the the ratios $T$.
\label{tab:CR}}
\end{table}

\subsection{General Patterns: $\epe$}
Let us elaborate on our results for $\epe$.
The dominant supersymmetric contributions to $\eps$ and $(\Delta M)_{d,s}$
originate in the functions $S_H(t,t)$ and $S_\chi(t,t)$ contributing to
$F_{tt}$ in (\ref{tt}). These two functions add positive contributions
to $F_{tt}$ which is positive in the SM.  
As a result, the values of $\imlt$ and $|\relt|$ in the MSSM are smaller 
than in the SM for any choice of the supersymmetric parameters 
in (\ref{susypar}). This is clearly seen in table~\ref{tab:R}.
The suppressions of $\imlt$ and $|\relt|$ are largest for a very light,
mainly right-handed stop, for a large splitting between the stops, for
small $\tilde m_{\chi_1}$ and small $\tan\beta$.

The dominant supersymmetric contributions to $F_{\varepsilon'}$ in
(\ref{FE}) originate in supersymmetric contributions to the function
$Z$ in (\ref{eq:3b}) which results from $Z^0$-penguin diagrams and
photon-penguin diagrams. The corresponding contributions to the
functions $X$, $Y$, and $E$ have a considerably smaller impact on
$F_{\varepsilon'}$. This is related to a large extent to the fact that
the coefficient $P_Z$ in (\ref{FE}) is substantially larger than the
coefficients $P_X$, $P_Y$ and $P_E$.

Setting the parameters of tables 
\ref{tab:pbendr} and \ref{tab:inputparams} to their central values
we find
\be\label{fep}
F_{\varepsilon'}=10.3-7.7\cdot Z +1.1
\ee
where the last term represents the contributions of $X$, $Y$
 and $E$ evaluated within the SM.
The first term represents the part unaffected by supersymmetric
contributions. It is dominated by the contribution of the QCD-penguin
operators, whose Wilson coefficients are strongly enhanced by
renormalization group evolution from scales $\ord(\mw)$ to scales
$\ord(1~\gev)$. As stated above, the supersymmetric contributions
affect $F_{\varepsilon'}$ dominantly through the function $Z$, which 
equals 0.66 in the SM.

The charged Higgs contribution to the function $Z$ is positive for
any value of $\tan\beta$ and $m_{H^\pm}$. Consequently, as seen
in (\ref{fep}), this contribution suppresses $F_{\varepsilon'}$
and $\epe$. This suppression of $\epe$ is enhanced by the
suppression of $\imlt$ through charged Higgs contributions as
discussed above. A detailed numerical analysis of these
effects within the two Higgs doublet model has been presented
in \cite{EP99,BBHLS} 

As pointed out in \cite{GG}, depending on the choice of
the supersymmetric parameters, the chargino contribution
to the function $Z$ and more generally to $F_{\varepsilon'}$
can have either sign. Indeed as seen in table~\ref{tab:K}
the function Z can be suppressed up to $30\%$ relative to
its SM value even if all constraints are imposed.
If this happens the suppression of $\epe$ due to $Z^0$-penguins 
observed in the SM and in the two-Higgs doublet model 
becomes smaller. 

\begin{figure}[tb]
  \begin{center}
     \epsfxsize=11.5truecm
    \begin{turn}{-90}
    \epsffile{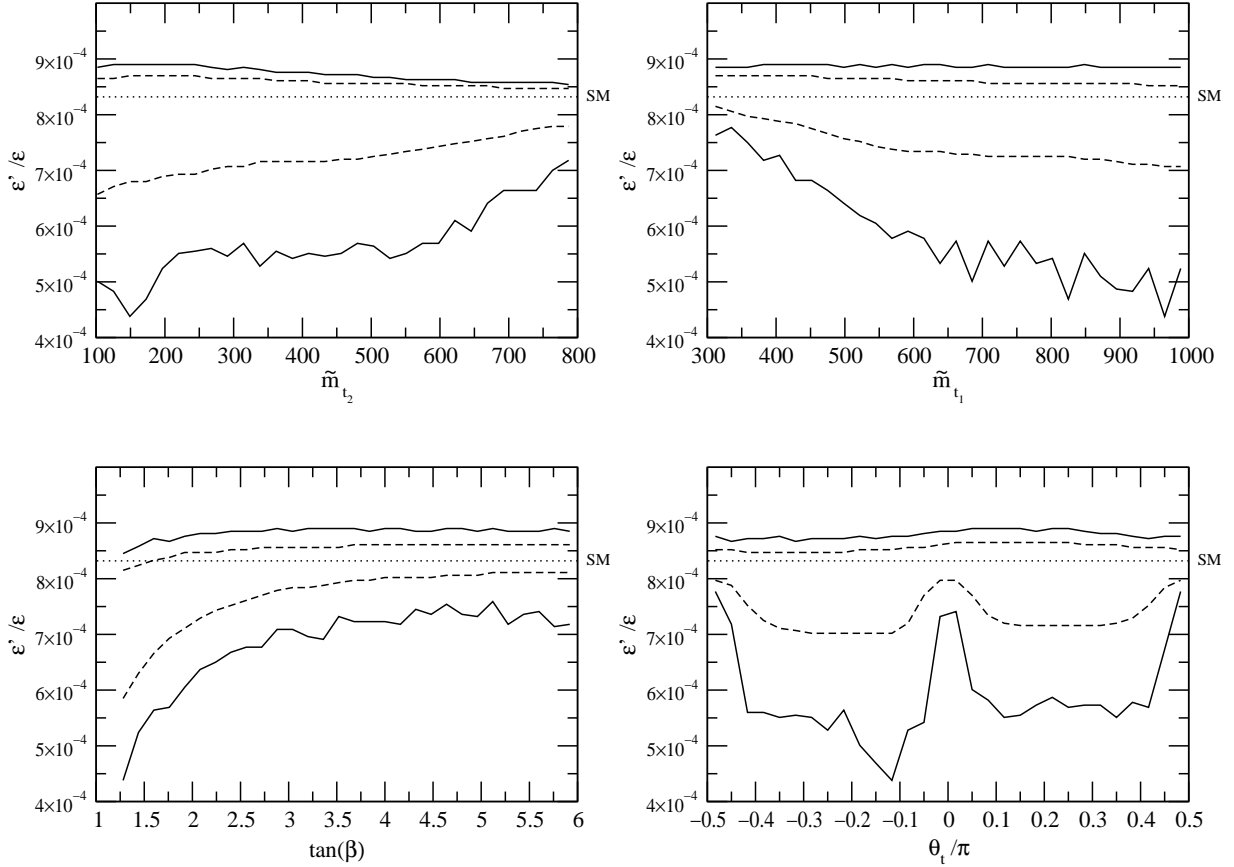}
    \end{turn}   
    \caption{Contour plots of $\epe$ as a function of
      $\tilde m_{t_2}$, $\tilde m_{t_1}$, $\tan \beta$ and
      $\theta_{\tilde t}$, imposing the constraints discussed in
      section \ref{sec:constraints}. The dashed and continuous curves
      contain $95\%$ and $100\%$ of the allowed points respectively.
      The dashed horizontal line corresponds to the SM prediction for
      central values of the parameters in table
      \ref{tab:inputparams}.}
    \label{fig:scatter_epp}
  \end{center}
\end{figure}

Whether $\epe$ can be enhanced in MSSM over the SM prediction depends
on whether the positive contribution of charginos to the function
$F_{\varepsilon'}$ can overcompensate the corresponding negative
charged Higgs contribution to this function and the suppression of
$\imlt$ due to supersymmetric contributions. While the chargino
contribution to $F_{\varepsilon'}$ can indeed overcompensate the
corresponding negative charged Higgs contribution (see
Table~\ref{tab:K}) the resulting enhancement of $F_{\varepsilon'}$ is
generally insufficient to cancel the suppression of $\imlt$.
Consequently for the dominant part of the allowed values of
supersymmetric parameters $\epe$ in MSSM is suppressed with respect to
its SM estimate. This is clearly seen in the contour plots in
fig.~\ref{fig:scatter_epp} where we show the range for $\epe$ as a
function of $\tilde m_{t_2}$, $\tilde m_{t_1}$, $\tan \beta$ and
$\theta_{\tilde t}$, with other parameters varied in the allowed
ranges.  The strongest suppression of $\epe$ takes place for
lowest $\tan\beta$, $m_{H^\pm}$, for a light, mainly right-handed stop
and for large splitting between the stops. The enhancement of $\epe$
over its SM value is found for large $\tan\beta$ and $m_{H^\pm}$ and
small $m_{\chi_1}$. In this case the charged Higgs contribution
becomes much smaller than the chargino contribution and the
enhancement of $F_{\varepsilon'}$ through the chargino contribution can
overcompensate the suppression of $\imlt$. 

Finally, let us comment on the dependence of the above results on the
hadronic parameters entering $\epe$. It is well known that for central
values of the parameters in table \ref{tab:inputparams} there is a
strong cancellation between QCD and electroweak penguin
contributions, as shown explicitly in eq.~(\ref{fep}). In this case,
the enhancement or suppression of electroweak penguins due to SUSY
contributions has a relatively large impact on the value of $\epe$.
If, however, $B_6^{(1/2)}$ is larger than one, $\epe$ becomes more
QCD-penguin dominated, the cancellation becomes much less effective
and therefore also SUSY contributions have a smaller impact on the
value of $\epe$ (as discussed above, the main contribution to QCD
penguins at the hadronic scale comes from the mixing with
current-current operators in the RGE evolution, and is therefore
independent of SUSY contributions). This is clearly displayed in
table \ref{tab:epe_had}, where we report the extremal values of
$K(F_{\varepsilon^\prime})$ and $T(\epe)$ for $B_6^{(1/2)}=1$, $1.5$ and 
$2$. Similarly, increasing and decreasing the value of $B_8^{(3/2)}$,
the effects of SUSY contributions to $\epe$ are increased and
decreased respectively.

\begin{table}[t]
  \begin{center}
\begin{tabular}{|c|c|c|c|c|c|}
\hline
& & \multicolumn{2}{c|}{{\rm No Constraints}} &
  \multicolumn{2}{c|}{{\rm All Constraints}} \\
\hline
$B_6^{(1/2)}$ & $T $, $K$ & {\bf min} & {\bf max} & {\bf min} & {\bf max} \\
\hline
$1.0$ & $K(F_{\varepsilon^\prime})$ & $0.59$ & $1.27$ & $0.76$ & $1.23$    \\
$1.0$ & $T(\epe)$ & $0.42$ & $1.07$ & $0.53$ & $1.07$     \\
$1.5$ & $K(F_{\varepsilon^\prime})$ & $0.78$ & $1.13$ & $0.87$ & $1.11$    \\
$1.5$ & $T(\epe)$ & $0.53$ & $1.02$ & $0.60$ & $1.02$     \\
$2.0$ & $K(F_{\varepsilon^\prime})$ & $0.85$ & $1.08$ & $0.91$ & $1.07$    \\
$2.0$ & $T(\epe)$ & $0.55$ & $1.00$ & $0.63$ & $1.00$     \\
\hline
\end{tabular}
    \caption{The minimal and maximal values of the the ratios
      $K(F_{\varepsilon^\prime})$ and $T(\epe)$ for different values
      of $B_6^{(1/2)}$,
      without constraints and with all constraints taken into account.}
    \label{tab:epe_had}
  \end{center}
\end{table}

\subsection{Global Analysis}
\label{sec:global-analysis}

In principle we could next vary all the SM parameters of tables
\ref{tab:pbendr} and \ref{tab:inputparams} in order to obtain the full
ranges for $\epe$ and for the branching ratios of rare decays in the
MSSM. We do not think such an analysis would be very instructive at
present. Roughly speaking the full ranges for the quantities
considered in this paper can be obtained by using the SM ranges, given
for instance in \cite{AJBLAKE}, and rescaling them by the enhancement
and suppression factors of table \ref{tab:R}.

On the other hand we have updated our 1999 analysis of
$(\epe)_{\mathrm SM}$ \cite{EP99} by modifying simply $\OEE=0.25$ to
$\OEE=0.16$ recently obtained in \cite{ECKER99}. The result in NDR is
given in table \ref{tab:31738}, to be compared with
$\epe=(7.7^{+6.0}_{-3.5})$ of our previous analysis.

\section{Summary}
\label{sec:summary}
\setcounter{equation}{0}
We have analyzed the CP violating ratio $\epe$ and rare K and B decays in the 
MSSM with minimal flavour and CP violation. A compendium of
phenomenological formulae that should be useful for future analyses is
given in section \ref{sec:compendium}.
In this version of the MSSM no new operators beyond those present in the SM 
contribute and
all flavour changing  transitions are governed by the CKM matrix.
In particular there are no new phases beyond the CKM phase. 
This implies that the supersymmetric effects enter various quantities
of interest exclusively through the one-loop functions.

In spite of this simplification the analysis is complicated by the
fact that in addition to a number of poorly constrained supersymmetric
parameters one has to deal with hadronic uncertainties. The latter
plague in particular $\epe$ and to a lesser extent $\varepsilon$ and
$B_{d,s}^0-\bar B_{d,s}^0$ mixings.  On top of this there are
uncertainties in the CKM parameters. Therefore in order to uncover
supersymmetric effects we have set the input parameters like $\vert
V_{ub}\vert$, $\vcb$, $\hat B_K$, $\bsi$, $\bei$, that are insensitive
to supersymmetric parameters, to their central values.  Imposing
subsequently constraints on the supersymmetric parameters coming from
$\varepsilon$, $B_{d,s}^0-\bar B_{d,s}^0$ mixings, $B\to X_s \gamma$,
$\Delta\varrho$ in the electroweak precision studies and from the
lower bound on the neutral Higgs mass, we have calculated the MSSM
predictions for $\epe$ and rare branching ratios normalized to the SM
results.  Our findings are summarized in table \ref{tab:R} and in the
abstract. The calculated quantities in the K-system are generally
suppressed by supersymmetric effects, although a $7\%$ enhancement of
$\epe$ is still possible. In the case of $\bnunu$ and $\bmumu$ both
suppressions and enhancements in the ballpark of $30-50\%$ are still
possible.  We have pointed out that an improved lower bound on the
neutral Higgs mass could considerably reduce the allowed ranges.

In view of very large hadronic uncertainties in $\epe$  it will be very
difficult to distinguish the MSSM prediction for this ratio from the one
 in the SM.
If future improved calculations of $\bsi$   and $\bei$   
will demonstrate indisputably that $(\epe)_{SM}$ is below the experimental 
data our results show that the MSSM will be of little help to remedy this
difficulty. On the other hand if it turns out that $(\epe)_{SM}$ is too 
large then supersymmetric effects in the MSSM could help to obtain 
the agreement with data.

As rare decays such as  $\kpnn$, $\klpn$, $\bnunu$ and $\bmumu$ 
are theoretically very clean, they offer a
much better place to look for supersymmetric effects within the MSSM than
$\epe$. The analyses of these decays will be particularly transparent as soon
as $(\Delta M)_d/(\Delta M)_s$ and the CP violating asymmetry in
 $B\to \psi K_S$  will be precisely measured allowing the construction of 
the universal unitarity triangle \cite{UUT}. If it turns out that the data
for $\kpnn$ and $\klpn$ are above the SM expectations, it will
become clear than more complicated supersymmetric extentions of the SM than
the constrained MSSM considered here are required to understand the 
experimental results. In the case of $\bnunu$ and $\bmumu$ no such
clear conclusions will be possible as both enhancements and suppressions
of these branching ratios are allowed. On the other hand with precise data
the MSSM predictions could in principle be distinguished from the SM ones.

\section*{Acknowledgments}
We would like to thank Manuel Drees for very informative discussions,
and specially for switching off our PC during the most important run
of this work.  This work has been supported in part by the German
Bundesministerium f\"ur Bildung and Forschung under the contracts 06
TM 874 and 05HT9WOA0. L.S. acknowledges the partial support by the
M.U.R.S.T. 

\section*{Appendix}
\setcounter{equation}{0}

Here we collect the functions which enter the basic functions
of section \ref{sec:basic-functions}. We use the notation of
\cite{GG,BBHLS}. 
\bea
S(x,y)&=&xy\left\{\left[\frac{1}{4}+\frac{3}{2}\frac{1}{1-x}
-\frac{3}{4}\frac{1}{(1-x)^2}\right]\frac{\log x}{x-y}
+\left(x\leftrightarrow y\right)
-\frac{3}{4}\frac{1}{(1-x)(1-y)}\right\}
\nonumber\\
L_1(x,y,z)&=&xy\left[F(x,y,z)+F(y,z,x)+F(z,x,y)\right]
\nonumber\\
L_2(x,y,z)&=&xy\left[xF(x,y,z)+yF(y,z,x)+zF(z,x,y)\right]
\nonumber\\
L_3(x,y,z)&=&\frac{1}{xy}L_2(x,y,z)
\nonumber\\
F(x,y,z)&=&\frac{x\log{x}}{(x-1)(x-y)(x-z)}
\nonumber
\eea
\vskip 1truecm
{\bf Box($\Delta$ S=1) }
\bea
B_{SM}(x)&=&-\frac{x}{4(x-1)}+\frac{x}{4(x-1)^2}\log{x}
\nonumber\\
B^{u}_{\chi}(x,y,z)&=&-\frac{1}{xy}L_1(x,y,z)
\nonumber\\
B^{d}_{\chi}(x,y,z)&=&L_3(x,y,z)
\nonumber
\eea
\vskip 1truecm
{\bf $\gamma$-penguin }
\bea
D_{SM}(x)&=&\frac{x^2\left(25-19x\right)}{36\left(x-1\right)^3}+
\frac{\left(-3x^4+30x^3-54x^2+32x-8\right)}{18\left(x-1\right)^4}
\log{x}
\nonumber\\
D_{H}(x)&=&\frac{x\left(47x^2-79x+38\right)}{108\left(x-1\right)^3}+
\frac{x\left(-3x^3+6x-4\right)}{18\left(x-1\right)^4}
\log{x}
\nonumber\\
D_{\chi}(x)&=&\frac{\left(-43x^2+101x-52\right)}{108\left(x-1\right)^3}+
\frac{\left(2x^3-9x+6\right)}{18\left(x-1\right)^4}
\log{x}
\nonumber
\eea
\vskip 1truecm
{\bf $Z^0$-penguin }
\bea
C_{SM}(x)&=&\frac{x\left(x-6\right)}{8\left(x-1\right)}+
\frac{x\left(3x+2\right)}{8\left(x-1\right)^2}
\log{x}
\nonumber\\
C_{H}(x)&=& \frac{x}{8(x-1)}-\frac{x}{8(x-1)^2}\log{x}
\nonumber\\
C_{\chi}^{(1)}(x,y)&=&\frac{1}{16\left(y-x\right)}\left[\frac{x^2}{x-1}\log{x}
-\frac{y^2}{y-1}\log{y}\right]
\nonumber\\
C_{\chi}^{(2)}(x,y)&=&\frac{\sqrt{xy}}{8\left(y-x\right)}\left[
\frac{x}{x-1}\log{x}-\frac{y}{y-1}\log{y}\right]
\nonumber
\eea
\vskip 1truecm
{\bf Gluon-penguin }
\bea
E_{SM}(x)&=&\frac{x\left(x^2+11x-18\right)}{12\left(x-1\right)^3}+
\frac{\left(-9x^2+16x-4\right)}{6\left(x-1\right)^4}
\log{x}
\nonumber\\
E_{H}(x)&=&\frac{x\left(7x^2-29x+16\right)}{36\left(x-1\right)^3}+
\frac{x\left(3x-2\right)}{6\left(x-1\right)^4}
\log{x}
\nonumber\\
E_{\chi}(x)&=&\frac{\left(-11x^2+7x-2\right)}{36\left(x-1\right)^3}+
\frac{x^3}{6\left(x-1\right)^4}
\log{x}
\nonumber
\eea

\end{document}